# Dosimetry for FLASH Radiotherapy: A Review of Tools and the Role of Radioluminescence and Cherenkov Emission

**Muhammad Ramish Ashraf\*[1], Mahbubur Rahman\*[1], Rongxiao Zhang[1,2,3], Benjamin B. Williams[1,2,3], David J. Gladstone[1,2,3], Brian W. Pogue[1,3], Petr Bruza[1]**

[1] Thayer School of Engineering, Dartmouth College Hanover NH 03755, US

[2] Department of Medicine, Geisel School of Medicine, Dartmouth College Hanover NH 03755, USA

[3] Norris Cotton Cancer Center, Dartmouth-Hitchcock Medical Center, Lebanon, NH 03756, USA

\* Authors contributed equally.

Corresponding Authors: Petr Bruza (Petr.Bruza@dartmouth.edu)

Copyright Information[1]

**Abstract**

While spatial dose conformity delivered to a target volume has been pushed to its practical limits with advanced treatment planning and delivery, investigations in novel temporal dose delivery are unfolding new mechanisms. Recent advances in ultra-high dose radiotherapy, abbreviated as FLASH, indicate the potential for reduction in healthy tissue damage while preserving tumor control. FLASH therapy relies on very high dose rate of > 40Gy/sec with sub-second temporal beam modulation, taking a seemingly opposite direction from the conventional paradigm of fractionated therapy. FLASH brings unique challenges to dosimetry, beam control, and verification, as well as complexity of radiobiological effective dose through altered tissue response. In this review, we compare the dosimetric methods capable of operating under high dose rate environments. Due to excellent dose-rate independence, superior spatial (~<1 mm) and temporal (~ns) resolution achievable with Cherenkov and scintillation-based detectors, we show that luminescent detectors have a key role to play in the development of FLASH-RT, as the field rapidly progresses towards clinical adaptation. Additionally, we show that the unique ability of certain luminescence-based methods to provide tumor oxygenation maps in real-time with submillimeter resolution can elucidate the radiobiological mechanisms behind the FLASH effect. In particular, such techniques will be crucial for understanding the role of oxygen in mediating the FLASH effect.

---

## 1. Introduction:

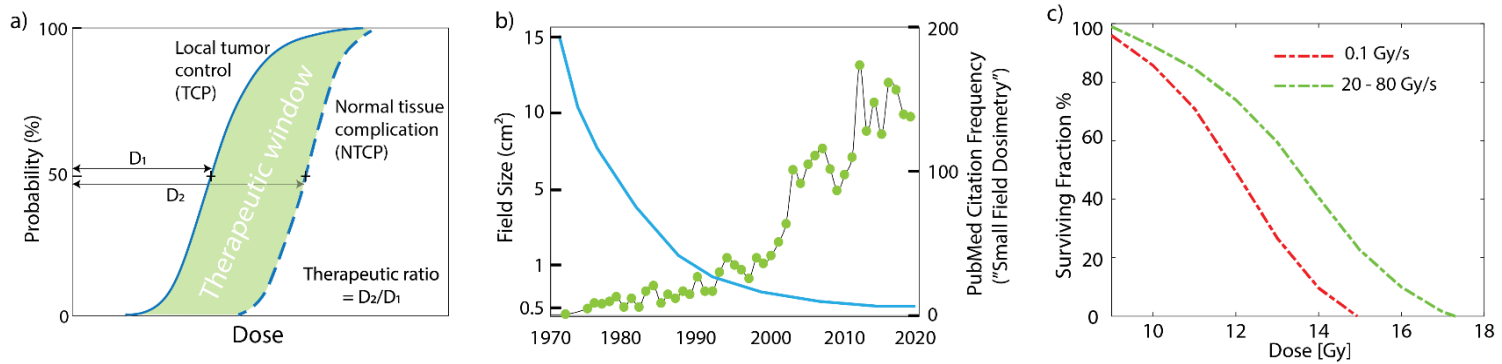

Figure 1) **a)** Typical dose-response curves in radiotherapy; maximum tumor control with minimal normal tissue complication is desired. **b)** Over last the few decades this has been possible due to spatial modulation of the beam, leading to an increasing use of small fields (<10 mm). Unfortunately, this led to complex dosimetric issues unique to small fields that standard dosimeters were not suitable for. The x-axis in the in b) denotes the time-line of major advancements that have happened in the field of external beam radiotherapy. Left scale denotes how typical field sizes have varied with these advancements and right scale shows the number of peer-reviewed publications per year based on a PubMed search of the phrase "Small Field Dosimetry". **c)** The effect of ultra-high dose rates on cell survival curves. Data adapted from Hornsey et al[1], where decreased cell killing was seen in mouse intestine at high dose-rates versus low dose-rates. The decrease in cell-killing was attributed to rapid depletion of oxygen, which is required to 'fix' DNA damage, at high dose-rates.

Decades of research in radiation therapy has been focused towards increasing the therapeutic ratio[2] (Figure 1a), and many techniques such as inverse treatment planning optimization, intensity modulated radiation therapy, or on-board imaging guidance have achieved this goal primarily via higher spatial modulation of the primary beam (Figure 1b). Temporal modulation of dose has also been widely exploited in its relation to repair of sublethal damage and the cell cycle, and this has been widely adopted in almost all clinical treatments through fractionated treatment plans. Yet, the role of higher dose-rate effects has been largely undeveloped in clinical treatment[3]. Interesting early studies[4–8] (1960-1980) had observed peculiar effects of reduced cell killing at ultra-high dose-rates, such as the study by Hornsey et al[1], who illustrated reduced cell killing in mouse intestine at high dose rates(Figure 1c). However, a recent study (2014) by Favaudon et al[9] has sparked explosive interest in ultra-high dose rates again. Contrary to conventional radiotherapy techniques, which employ mean dose-rates of ~0.03 Gy/s with doses of ~2 Gy delivered over 10-30 fractions, the authors used an ultra-high mean-dose rate of 40 Gy/s with total irradiation time <500 ms to achieve an improved differential response between tumor and normal tissue. They reported less normal tissue damage with high dose-rate irradiation when compared to 'conventional' radiotherapy conditions, while observing similar anti-tumor response in both modalities. The authors termed this phenomenon as the 'FLASH' effect. Multiple groups have now replicated the FLASH effect in different murine organs[10,11] and in superficial treatments in animal models such as mini-pigs, cats[12] and zebrafish[13]. It has been shown that the FLASH effect can be triggered using electrons[9,14,15], x-rays[16,17] and protons[18,19]. Recently, the first human patient was treated with FLASH-RT[18], with promising clearance of the lymphoma lesion with lower skin toxicity than has been seen in previous irradiations. The field of FLASH-RT has experienced wide interest and growth and could be a critical area of development for better normal tissue sparing in radiotherapy.

The calibration and quality assurance tools for dosimetry also need to be adapted accordingly to keep up with the ever-changing nature of radiation dose delivery Historically, spatial modulation in dose delivery has led to increased use of small radiation fields or beamlets for which accurate dosimetric characterization was found to be non-trivial. These problems associated with small fields are well documented[20]. This warranted a need for high-resolution detectors which most vendors now typically provide for the measurement of cumulative dose distributions[21]. With the emergence of FLASH-RT and other promising high dose-rate modalities, such as Microbeam Radiation Therapy (MRT)[22] and Synchrotron stereotactic radiotherapy[23], it is expected that new dosimetric challenges will arise. Spatiotemporal dosimetry for small radiation beamlets delivered dynamically under high dose-rate conditions can be difficult. For successful translation of FLASH-RT to a clinical setting, dosimetry must be performed accurately and rigorously keeping in mind limitations of various detectors at high dose-rates and non-standard nature of the various FLASH irradiation platforms. The issue of dosimetric uncertainty in preclinical radiobiological studies and its effect on reproducibility and eventual translation to clinics has been highlighted by multiple authors[24,25]. The National Institute of Standards and Technology (NIST) has recommendations regarding accurate measurement and reporting of dose in preclinical radiobiological studies[26]. In a multi-institution audit of irradiator output[25], it was found that only one facility was able to deliver a dose within 5% of the prescribed limit; other facilities had errors ranging from 12% to 42%. The issue is primarily because of the non-standard nature of irradiation platforms used in pre-clinical studies and a lack of protocols and guidance. To tackle this issue, the American Association of Physicists in Medicine (AAPM) has initiated Task Group No. 319 "Guidelines for accurate dosimetry in radiation biology experiments". One of the major aims of the task group is to standardize dosimetry and review uncertainties associated with non-clinical units used in preclinical radiobiological studies. Therefore, this review is motivated by the need to assess the various different uncertainties associated with performing accurate dosimetry under FLASH and high dose-rate irradiation conditions.

To this end, dosimetric problems unique to high dose-rate environment such as FLASH will be identified in this review. Common radiation dosimeters based on different physical principles, mainly, *luminescent*, *charged* based and *chemical* detectors will be discussed in light of the dosimetric issues identified earlier. We hypothesize that based on the underlying physical mechanisms, luminescent based detectors can be used to perform accurate, real-time dosimetry, with predetermined corrections under reference conditions to account for non-ideal characteristics (e.g. quenching of scintillator response for high LET particles and energy dependence of Cherenkov radiation). It can be argued that the most important dosimetric aspects of FLASH-RT (discussed in detail in section 2) are dose-rate independence, spatial resolution, and temporal resolution of a detector. Comparing these three characteristics of the detectors, their typical usage, and the underlying physics, it can be predicted that luminescent detectors offer unparalleled spatial-temporal resolution and dose-rate independence (Figure 2). Values in Figure 2 are based on typical usage and exceptions to these do exist; these exceptions will be noted, where appropriate, in the text. Another purpose of this review is to provide an overview of tools that have been used in high-dose rate conditions. Therefore, instances of different dosimetric tools used in FLASH-RT and other high dose-rate modalities will be consolidated and summarized. Finally, the mediating role of oxygen tension in FLASH-RT will be discussed briefly. The superior ability of luminescent based methods to sense oxygen tension and measure dose and LET (simultaneously, specifically for particle therapy) in real-time will be described for its potential in understanding the radiobiological mechanisms underlying the protective effect of FLASH on normal tissue.

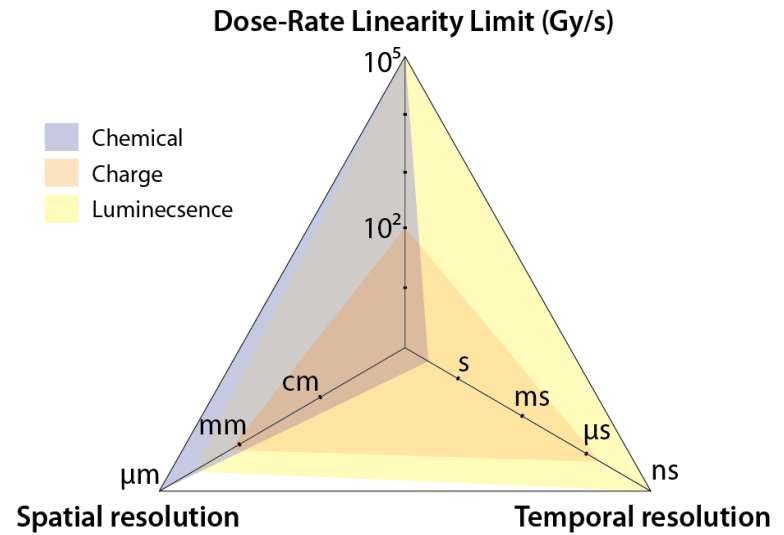

Figure 2) Spider plot comparing the three different categories of detector. The axes denote major dosimetric issues associated with FLASH-RT. It can be seen that luminescent detectors can provide ~ns time-resolution, with sub-millimeter resolution and dose-rate independence up to a dose rate of $10^5$ Gy/s.

## 2. Dosimetric Aspects of FLASH-RT That Need to be Considered

### 2.1 Dose-Rate Dependency

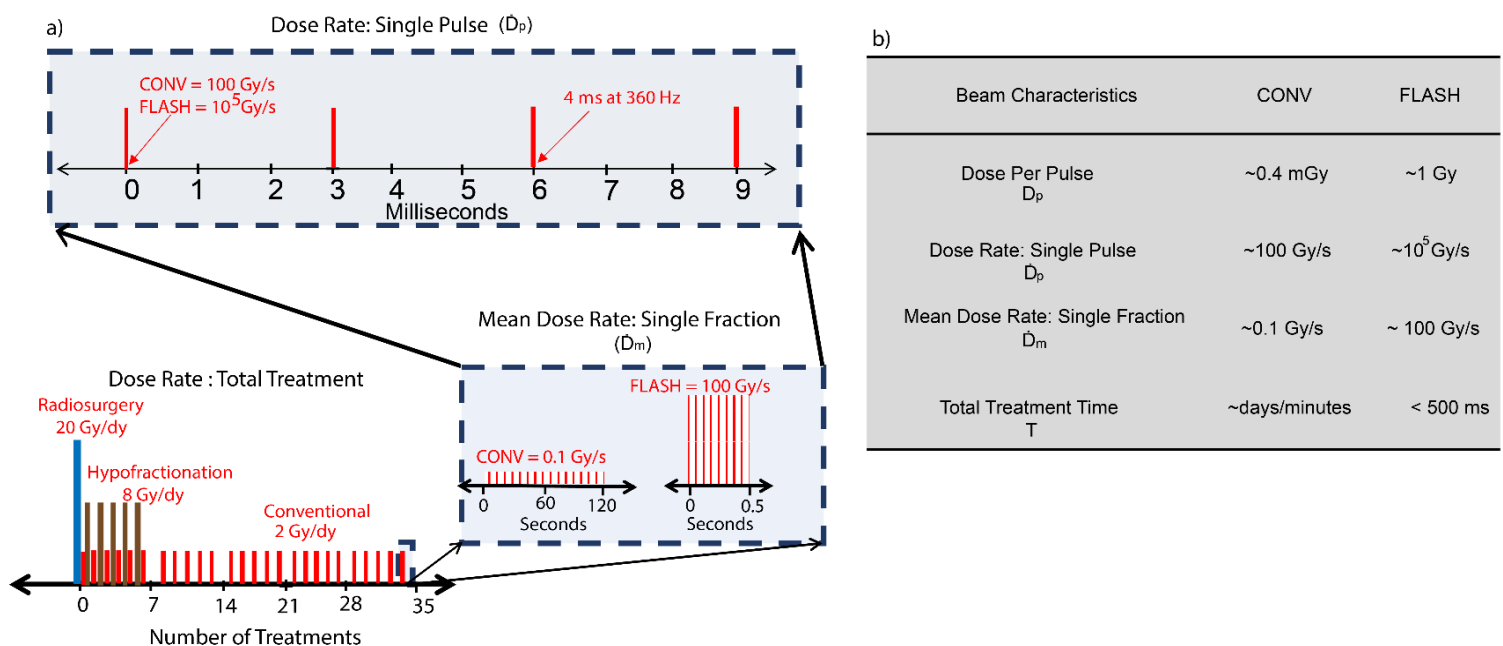

Fig. 3) a) Dose rate schemes in radiation therapy indicating different interpretations of dose-rate. b) Typical temporal beam characteristics for conventional (CONV) and FLASH-RT using electrons.

It is instructive to first compare temporal structure of FLASH and conventional radiotherapy beams in order to understand problems unique to high dose-rate conditions. Typically, radiation sources used in radiotherapy are not continuous but rather pulsatile in nature. The repetition rate and duty cycle depend on the type of particle acceleration used. For instance, most clinical linear accelerators (linac) typically have a pulse duration of 3-5 μs with repetition rate of ~200-400 Hz. For cyclotron based proton beams, the beam can be considered to be quasi-continuous due to short pulse duration and repetition rates on the order of a few nanoseconds. Additionally, modern radiation techniques typically deliver dose in fractionated manner over a period of few days. Therefore, dose-rate can be either the defined over the course of a whole treatment, one fraction or within a single pulse. The different interpretations of dose-rate are illustrated in Figure 3a. Table in Figure 3b presents a side by side comparison of temporal beam characteristics of FLASH and CONV radiation therapy; the dose per fraction is denoted as the mean dose-rate, $\dot{D}_m$. The dose-rate in a pulse or the instantaneous dose-rate is denoted as $\dot{D}_p$, which is the ratio of dose delivered in a pulse ($D_p$) divided by the pulse duration. This is an important distinction because instantaneous dose-rates in conventional radiotherapy can be comparable or even higher when compared to the mean dose-rate ($\dot{D}_m$) of 40 Gy/s used to trigger the FLASH effect in current preclinical studies. Note that the conventional beam characteristics are based on typical linac based clinical beams and the values for FLASH are based on prototype electron linacs that have been successfully used to elicit the FLASH effect[27]. Multiple studies have now rigorously explored the beam parameters needed to trigger a reproducible FLASH effect and it has been shown that the $\dot{D}_p$ and $D_p$ play a critical role[27,28]. For FLASH-RT, these quantities can be orders of magnitude higher as shown in 3b), leading to issues of saturation, and non-linear response of standard dosimeters at large doses.

### 2.2 Spatial Resolution

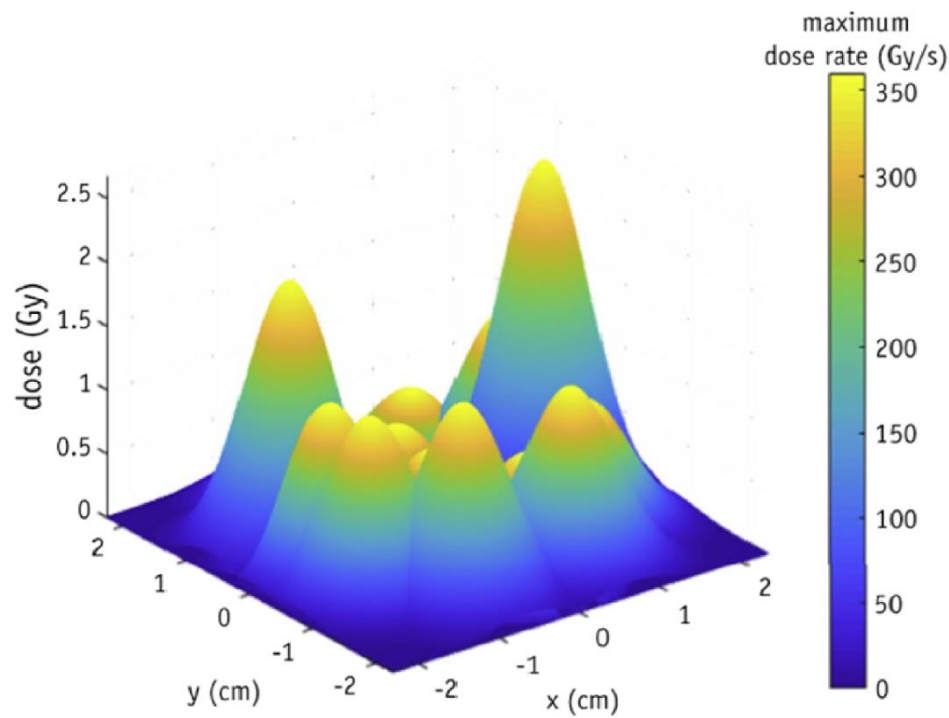

Figure 4) Spatial distribution of dose-rate within a pencil beam scanned proton beam illustrating the role of scatter contribution from adjacent spots, which leads to an inhomogeneous distribution of dose-rate. A point detector or an imaging array with coarse detector spacing will be unable to measure dose-rate distribution with accuracy. Figure adapted from Marlen et al[29]. Reprinted from "Bringing FLASH to the Clinic: Treatment Planning Considerations for Ultra high Dose-Rate Proton Beams", Vol 106 / Issue 3, Figure 2A, Copyright (2020), with permission from Elsevier.

Many preclinical FLASH studies have been performed on small animal models and tumor volumes. As mentioned earlier, performing accurate dosimetry with small beamlets is notoriously difficult. If the size of the sensitive volume of the detector is comparable to the radiation field size, dose averaging effects can lead to erroneous measurement of absorbed dose and artificial broadening of the field penumbra. Multiple preclinical studies have been performed using the experimental Oriatron eRT6- 6MeV and Kinetron 4.5 MeV linear accelerator (PMB-Alcen, Peynier, France). Design limitations of these linacs result in only small field openings. Small applicators used to confine the radiation field to the organs of interest can also introduce additional complex problems[30]. Additionally, clinical linacs modified to deliver flash dose rates[31] tends to utilize significantly shorter source to target distance to further increase the dose rate. Close proximity to source renders only small field sizes viable. Varian's Clinac 21EX was modified by Schüler et al[31] and it was found that mean dose-rate of 900 Gy/s was achievable near the transmission ionization chamber in the gantry head, where 90% of the field diameter was measured to be ~12 mm. For the beam energy used in the study (20 MeV), this falls in the realm of small field dosimetry, due to the lateral electronic disequilibrium. While small field dosimetry issues are not specific to FLASH, they nonetheless contribute to dosimetric complexity.

A related issue specific to FLASH is that of spatial averaging of dose-rate. Accurate determination of fully spatially-resolved dose-rate is critical for FLASH accuracy. This is even more important currently, since one of the main goals is to optimize the beam parameters so that a FLASH effect can be elicited in the clinical setting. A simulation study conducted by Marlen et al[29] explored the idea of spatial dose-rate distribution within a broad beam and its effect on FLASH response. Their study was based on pencil beam scanning with protons, but the results can be generalized to other modalities. Large field sizes are obtainable with protons when spot scanning techniques are used. However as reported in the study, dose rate in one spot will be affected by the low dose-rate scatter contributed by the adjacent spots (shown in Figure 4). They reported that only 40 % of the dose is delivered at FLASH rates for a spot peak dose rate at the center of 100 Gy/s. When the peak dose rate is increased to 360Gy/s, the contribution to FLASH dose rates increases to 75%. Rahman et al. also measured, using a scintillating sheet, as much as 41% standard deviation in maximum dose rate distribution for spot spacing as large as 10 mm (shown in figure 10 a)[32]. This will also hold true for broad electron beams, synchrotron produced x-ray beams and passively scattered proton beams, where scatter can contribute to an inhomogeneous distribution of dose-rates across the irradiated volume. This phenomena was also studied by Van de Water et al[33] where they investigated the possibility of achieving FLASH dose-rates with conventional proton pencil beam scanning and intensity modulated proton therapy techniques. They proposed a new metric to quantify spatially varying dose-rate in three dimensions, dose averaged dose-rate (DADR) and dose-weighted average of instantaneous dose-rate of all spots. It remains to be seen if such a criterion has any potential clinical value, but it points towards the need of a high resolution imaging detector. Additionally, as FLASH-RT continues to move towards clinical translation, large field sizes will need to be delivered at high dose-rates. Spatial distribution of dose and dose-rate can only be reliably measured using imaging techniques or detector arrays. Thus, a high resolution detector array with small inter-detector spacing is needed, so that a 2D spatial distribution of dose-rate can be measured accurately, avoiding any volume averaging effects. A 3D distribution of dose-rate within patient geometry would be the ideal case.

### 2.3 Time Resolution

Another aspect of dosimetry under FLASH conditions is that of real-time dose monitoring vs passive dose monitoring. While dose-rate independence is an important requirement for FLASH-RT, the ability to verify machine output, dose delivered per pulse, and dose-rate in real-time is of considerable interest. For high dose-rates and dose per pulse conditions, real-time dose monitoring is non-trivial. Not only dose-rate independent dosimeters are required, but dosimeters with a high enough temporal resolution and high bandwidth read-out methods are needed. For instance, some excellent dose-rate independent detectors like radiochromic film, alanine and TLD's etc. can only provide passive dose monitoring. Additionally, even though certain dosimeters can provide online dose-monitoring, they encounter other issues at ultra-high dose-rates which limits their capability. For example, clinical linear accelerators employ a monitor chamber in the gantry head which records machine output in real-time and serves as a beam-off signal when the recorded dose matches the required dose. However, most commercial ionization chambers start to show saturation or decreased ion-collection efficiency at high dose per pulse conditions rendering online monitoring of dose problematic at FLASH dose-rates. It has been shown by Jorge et al[34] that despite correcting for ion-recombination effects at high doses per pulse, ICs can show deviations up to 15%.

### 2.4 Dynamic Range

Dynamic range of a dosimeter is another issue pertinent to dosimetry in FLASH-RT. It has been shown that the oxygen depletion effect is highly dependent on the total dose delivered[35-37], with oxygen being depleted rapidly at high doses. If the oxygen depletion effect is indeed the underlying mechanism of the normal-tissue protection, it would be expected that preclinical studies will move towards hypo- or single-fractionated regimen with high doses per fraction to increase the therapeutic ratio. Currently, FLASH studies have been performed where total doses up to 40 Gy have been delivered in a short duration. For in-vivo dose verification, this implies that a dosimeter is required which does not suffer from saturation effects and maintains a linear or otherwise predictable behavior in response to dose. Radiation damage can also potentially be an issue for sensitive detectors such as MOSFET, however in most cases, the damage threshold for dosimeters is orders of magnitude higher compared to the dose threshold where dosimeters start showing saturation and non-linearity.

### 2.5 Other Ideal Characteristics

While, the previous few sections were primarily focused on dosimetric aspects unique to FLASH-RT, there are still other characteristics that make certain radiation dosimeters ideal for dosimetry. These characteristics go hand in hand with the ones discussed previously because they will eventually lead to erroneous measurement of dose and subsequently dose-rate. Mainly, an ideal dosimeter should be tissue-equivalent and not perturb the radiation field so that it can serve as an accurate surrogate for dose measurement inside the patient. A parameter closely tied to tissue equivalence is the energy dependence of the dosimeter. Ideally, a dosimeter that is energy-independent is required, because one would want the detector to respond uniformly to irradiation irrespective of the radiation quality. For example, certain non-tissue equivalent detector can over respond when there is significant low energy scatter present in the radiation field leading to erroneous measurements. These issues are amplified at small fields and can cause significant errors. In general, correction factors for small fields can be divided in two categories: 1) volume averaging, 2) other non-ideal characteristics. While volume averaging is a purely geometrical concept and can be minimized by using dosimeters which are small in volume, non-ideal characteristics are harder to correct for. The volume averaging correction factor on the central axis is within 1-2% if the size of the detector is 1/4th of the diameter of the incoming beam[38]. For example, the  unshielded Stereotactic Field Diode (SFD) diode (0.6 mm diameter) by IBA has a volume averaging correction factor of 1.003 for a 5 mm circular beam[39]. However, the overall correction factor for the diode has been typically found to be < 1 in literature,  which implies that the diode tends to over respond to radiation[40]. This over-response is explained by the presence of high-Z silicon in the diode which offers an increased photoelectric absorption coefficient at lower energies.   In FLASH-RT, these non-ideal characteristics can be critical because of the use of small fields as discussed above.

### 3.  Radiation Dosimeters

#### 3.1 Charge Based Dosimeters

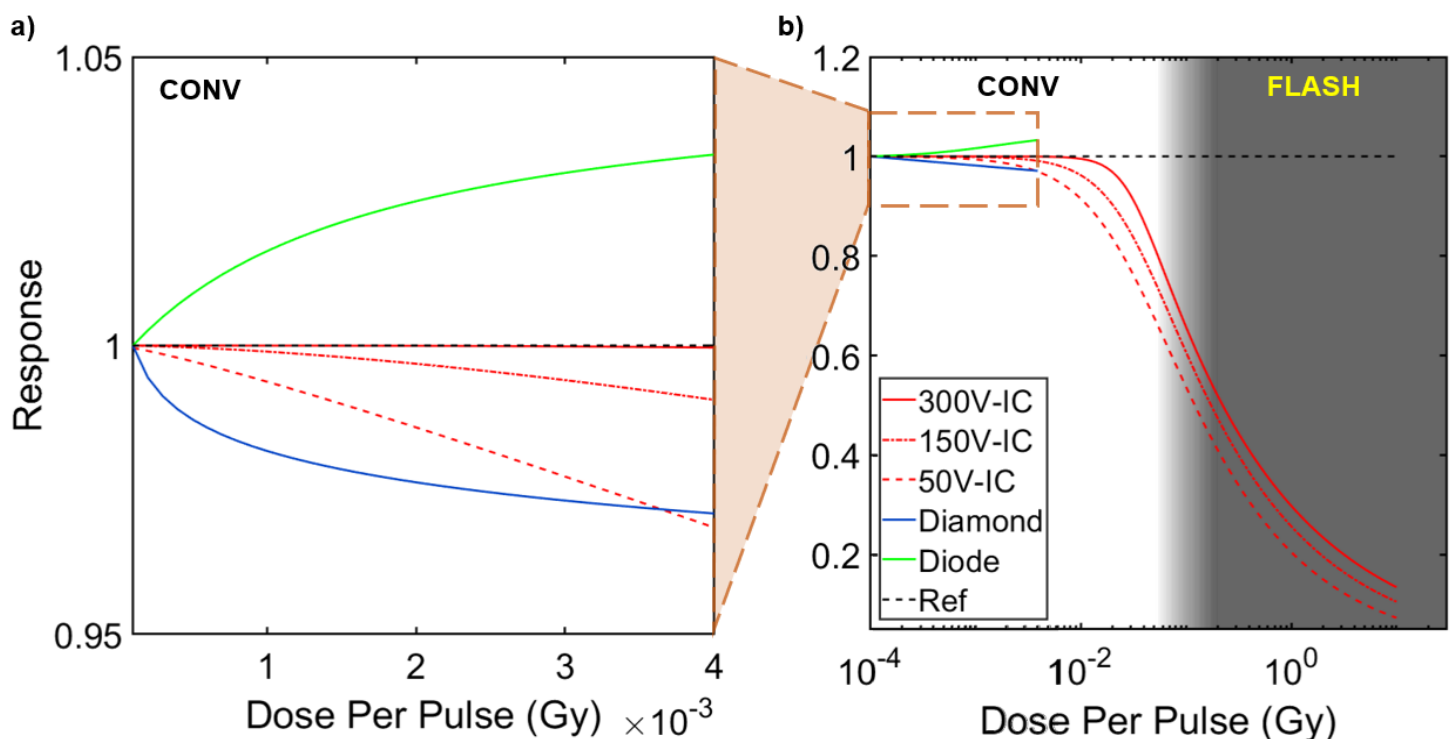

Fig 5) Model for charged based detector response based on Advanced Markus Chamber IC, PTW microDiamond, and Isorad Gold (n-type) diode detectors. Dose per pulse dependency of detector response are shown for a) conventional beams and b) FLASH beams. Advanced Markus Chamber IC response (charge collection efficiency) was calculated for three different bias voltages and the only charge-based detector to be tested in FLASH dose rates[41]. Model for diamond detector response (charge collection efficiency) and diode detector response (sensitivity) were only tested at conventional dose rates[42,43].

Most charged based dosimeters are based on the principle of creation of ion pairs or charges which can be collected and correlated to dose. In Ionization Chambers (IC), the collection of ion-pairs is facilitated by an application of an external bias across the electrodes. The voltage applied is typically high enough, such that all liberated charges are collected. However, at high doses per pulse, ion-pairs can recombine before they are swept across the E field and collected by the electrodes. This can lead to a decrease in sensitivity with increasing dose per pulse. While ion-recombination effects can be accounted for at moderate doses per pulse using Boag's model[44], at high dose per pulse the model breaks down. However, it should be noted that certain commercial IC chambers, such as the Advanced Markus IC by PTW can efficiently collect ions at dose-rates as high as ~ 300 Gy/s and a dose per pulse of ~5 mGy.

This is sufficient for conventional flattening filter free beams and pencil beam-scanned proton beams. Nevertheless, at high instantaneous dose-rates and doses per pulse typically used in FLASH, a correction factor would be needed to correct for ion-recombination. A study performed by Petersson et al[41], looked into ion-recombination effects of the Advanced Markus IC with their FLASH setup. The authors came up with a model to account for ion-recombination and polarity effect for the aforementioned chamber (Figure 5). It can be seen that the ion-collection efficiency has a strong dependence on dose per pulse above 1 Gy. With the correction factor applied, they concluded that the chamber could be used for FLASH dosimetry. In particular, they were able to measure a dose-per-pulse of 10 Gy. Interestingly, FLASH studies conducted with protons have employed ICs for dosimetric verification but without application of any correction factor. For example, Patriarca et al[18] employed a IC in their proton FLASH setup which used a cyclotron-based 230 MeV (IBA) proton beam as the radiation source. With cyclotron-based sources, high pulse repetition rates of around ~100 MHz can be achieved with pulse duration of around ~ 2ns. In this case, one can assume a 100% duty cycle, which drastically reduces the instantaneous dose rate or dose delivered in a pulse. Using Boag's method[44], the authors found that the total recombination of ions was around 1% at a maximum mean dose-rate of 80 Gy/s . Another study conducted by Beyreuther et al[13] used a 224 MeV proton beam at a dose-rate of 100 Gy/s and concluded that the pulse duration (~100 ms) was an order of magnitude higher than the ion-collection time for the Advanced Markus chamber (~10 us).

Solid-state detectors such as diamonds and diodes have been used extensively for dosimetry in modern radiotherapy techniques due to their high sensitivity and small size. The operation of silicon diodes and diamond detectors is similar to IC chambers in that radiation produces electron-hole pairs which can be collected. Essentially, they can be thought of as solid-state ICs. Whereas, direct recombination is the dominant process in IC, charge recombination in solid state detectors is a more complex process dominated by indirect recombination, because of the presence of RG (recombination-generation) centers and impurities that can act as trap centers. In general, dose-rate dependence of solid-state detectors can be modeled as $\sigma \sim D_r^{\Delta}$, where $\sigma$ is the electrical conductivity, $D_r$ represents the dose-rate and $\Delta$ is a fitting parameter which describes the dose-rate dependency of the detector[45]. Multiple studies have investigated the dose-rate dependence of diodes and diamonds. Ade et al[46] found the fitting parameter $\Delta$ for some diamond detectors to decrease by as much as 9% when the dose-rate was increased from 2.25 Gy/min to 3.07 Gy/min. Interestingly, diamond detectors have been have been reported to show increased or decreased sensitivity with increasing dose per pulse depending on their construction i.e. pure crystals, Chemical Vapor Deposition (CVD), high-pressure high-temperature HPHT[46]. While diamond detectors have not been extensively used for ultra-high dose-rate experiments, a microDiamond detector type 60019 (PTW) was used for one proof-of-concept proton-based FLASH study by Patriarca et al[18]. The sensitivity of the aforementioned diamond detector is shown in Figure 5. Recombination in diodes is also a complex physical process; whereas sensitivity of IC decreases with increasing dose per pulse, diodes are known to over respond at high dose per pulse[47–49]. The physical basis of this dependence is due to insufficient number of RG centers available for the excess minority carriers to recombine at high dose-rates and doses per pulse. Therefore, a larger fraction of charges is left behind and can be collected by the electrode, leading to increased sensitivity of the diode. To the best of our knowledge, no FLASH study has used diodes for dosimetric verification. Kinetic modelling of the recombination process in solid-state detectors has been carried out by multiple groups and we point the reader towards those references for a deeper understanding[42,48–50].

The time resolution of charged based dosimeters is mainly limited by the ion-drift velocity, mobilities of the different charge carriers present and other fundamental parameters such as the transit time (time taken for a charge to be completely collected) and minority carrier lifetime etc. For ICs with a typical external bias of 300 V, the temporal resolution usually ranges from a few ms to hundreds of ms[51]. For indirect band gap semiconductors, such as silicon, the time-resolution can be on the order of a few ms[52]. Pure diamond detectors, due to their superior electron and hole mobilities, can offer time resolution on the order of a few ns[53], whereas most synthetic diamonds ( i.e. CVD based) have a minority carrier lifetime of a few us[54]. Therefore, real-time dose monitoring is indeed possible with diodes, diamonds, and IC. However, it should be reiterated that the main limitation for such devices is the change in sensitivity, non-linearity, and saturation due to charge recombination at high instantaneous dose-rates relevant in FLASH-RT. For pre-clinical FLASH studies, multiple authors[19,55] have circumvented this issue by using a Faraday cup; a conductive metal cup which accumulates charge when put in the beam's path. The main advantage of using a Faraday cup is that saturation, ion-collection, and recombination effects can be avoided. Additionally, Faraday Cups have been used with nano-second time resolution in studies conducted with high energy and highly pulsed charged particle beams[56–58].

Typically, measurements performed with charge based detector are point (1D) or planar measurements (2D). The advantage of solid-state charge based detectors over gas-filled ICs is that they offer superior spatial resolution because of their increased sensitivity to radiation. For instance, the PTW microDiamond detector, if used in edge-on configuration (i.e. smallest dimension normal to the incident beam), exhibits a resolution of 1 μm. Another high resolution charge-based dosimeter of interest is the silicon single-strip detector (SSD) which has been touted as a potential dosimetric tool for Microbeam Radiation Therapy. The SSD was used at the European Synchrotron Radiation Facility and has been shown to demonstrate very high spatial resolution (~10 μm resolution) and high dynamic range.[59] Despite their high spatial resolution, one caveat of typical charge-based dosimeters is their energy dependence and tissue non-equivalence. This limits their usefulness in small field dosimetry which can lead to erroneous measurement of dose-rate in pre-clinical animal FLASH studies employing small beams.

### 3.2 Chemical Dosimeters

Certain materials that undergo structural changes, produce radicals, or change color when irradiated can be classified as chemical dosimeters. For example, when a solution of ferrous sulfate (Fricke) is irradiated, ferrous ions $Fe^{2+}$are oxidized to ferric ions, $Fe^{3+}$. The number of ferric ions produced is proportional to dose delivered and can be quantified by measuring the optical density of the solution. Fricke dosimeter was used by Hendry et al[55] in their study on effects of high dose-rate on oxygen concentration. However, diffusion of ferrous ions over time makes this technique sensitive to low dose-rates[60]. Similar in nature to the Fricke dosimeter, methyl viologen is another tool that was used by Favaudon et al[30] in their FLASH setup for online monitoring of dose. Dosimetry was performed by optical detection of the $MV^{\cdot+}$ radical at 603 nm. The authors were able to monitor dose synchronously with the electron-pulses, but a decay in the $MV^{\cdot+}$ radical with time (on the scale of a few minutes) was observed that ultimately led to loss in absorption; an issue which can play a major role at low dose per pulse/dose-rate conditions. Therefore, while, certain chemical dosimeters can provide absolute dosimetry and real-time detection of dose with ~ns resolution, the radiation induced species in these materials are generally not stable and can either diffuse spatially ($Fe^{3+}$) or decay with time($MV^{\cdot+}$) which makes such setups unsuitable for real-time dose monitoring in FLASH-RT.

Fortunately, chemical dosimeters that produce stable radiation-induced species are available. One such dosimeter is Alanine and has been extensively used in preclinical FLASH studies[12,34]. Alanine is an amino acid, which forms a stable free radical upon irradiation. The concentration of the free radical is proportional to the absorbed dose, which can be probed using an electron paramagnetic resonance (EPR) spectrometer. Alanine dosimeters exhibit a linear response over a large dynamic range (2Gy-150kGy) and are therefore routinely used in industrial facilities. Although, at doses below 2 Gy, alanine can show considerable relative uncertainty of ~1.5%[61]. However, this might not be an issue for FLASH dosimetry because generally high doses are needed to elicit the FLASH effect. The real value of alanine dosimeters for FLASH dosimetry is in their excellent dose-rate independence (up to ~3 x $10^{10}$ Gy/s[62]). Recently, alanine was used at the European Synchrotron Radiation Facility (ESRF), which is capable of producing really high dose-rates (~ 10kGy/s)[63]; the response of alanine was found to agree well with the PTW PinPoint IC, when the latter was corrected for ion-recombination effects.

Perhaps the major advantage of certain chemical based detectors in their inherent ability to provide planar or 3D measurements. One really popular and perhaps the major workhorse of the radiation dosimetry world, is the poly-diacetylene based self-developing radiochromic film. Upon irradiation, the film undergoes color change by polymerization. The change in color is typically quantified in terms of the optical density as measured by a densitometer or in some cases via microscopy. Radiochromic films can be considered to be the ideal dosimeter, in that they are energy independent, tissue equivalent, and demonstrate really high spatial resolution (sub-micron) limited only by the digitizing method. Additionally, dose-rate independence of radiochromic films is well established in literature and they have been found to be independent of any dose-rate effects up to a dose-rate of 15x $10^9$ Gy/s[64]. In fact, EBT3 Gafchromic Film (Ashland, Wilmington, DE) was used by Patriarca et al[18] to evaluate dose-rate independence of other dosimeters used in their FLASH setup. A detailed study was conducted by Jaccard et al[65(p3)] on the suitability of radiochromic film for high dose-rate FLASH dosimetry with the Oriatron eRT6 electron linear accelerator (PMB-Alcen, Peynier, France) and they concluded that film was independent up to a $\dot{D}_p$ of 8 x $10^6$ Gy/s. Additionally, radiochromic film has been used to measure dose homogeneity and verify field sizes in various different FLASH studies[9–11] and was also used to verify dose (along with alanine) for the human patient treated with FLASH-RT. One of the major drawbacks of radiochromic film is that measurements are performed offline, typically 24 hours post exposure to account for the fact that polymerization does not stop immediately after irradiation. In theory, real-time read-out of film can be performed with ms time resolution, since it has been reported in literature that polymerization is largely 'complete' within 2 ms of a 50 ns pulse[66]; however, as started earlier, polymerization still continues post-exposure which can act as a confounding variable for near real-time dosimetry. Nonetheless, attempts have been made at real-time readout of radiochromic film[67,68].

Certain chemical dosimeters can provide true 3D spatial dose distribution at high resolution and in patient geometry. This has been facilitated by the recent advent of gelatin-based polymers, which avoid the problem of ion diffusion encountered in Fricke dosimeters. However, diffusion in polymer gels can still occur in the first hour post irradiation and at high dose-gradients and high doses[69]. Essentially, polymer gels act as 3D radiochromic film, except that the change in optical density is probed in 3D as opposed to a single 2D plane. Multiple methods have been used to probe these radiation sensitive gels, such as MRI, x-ray CT, ultrasound and optical projection tomography (OPT)[69]. However, OPT has stood out as the more popular read-out method, because of the high spatial resolution it offers. Unfortunately, like all chemical dosimeters, some change in signal is expected with time. Change in signal post-exposure coupled with the fact that complicated read-out machinery is needed to probe the response to radiation, renders this technique unsuitable for real-time dose measurement. More importantly, in FLASH-RT context, polymer based gels have been known to show dose-rate dependence, which might be attributable to competing radiation induced chemical reactions in the gel and the dose-rate dependence of water radiolysis products[70]. The dose-rate dependence seems to be a function of concentration of oxygen scavengers in the matrix, with less dose-rate dependence seen at high concentrations of $O_2$ scavengers[71]. Dose-rate dependence is also a function of the type of monomer unit of the gel[70]. For a more detailed analysis on the origin of dose-rate dependence in polymer gels, we refer the reader to a comprehensive review by De Deene et al[70].

### 3.3 Luminescent Dosimeters

In this text, luminescence refers to any technique which utilizes generation of optical photons in response to radiation as a surrogate for dose. This generally includes thermoluminescent detectors (TLD), optically stimulated luminescence detectors (OSLD), organic/inorganic scintillators and Cherenkov radiation. Physical properties that enable luminescent detectors to be of value in FLASH-RT will be discussed.

#### 3.3.1) TLD and OSLD

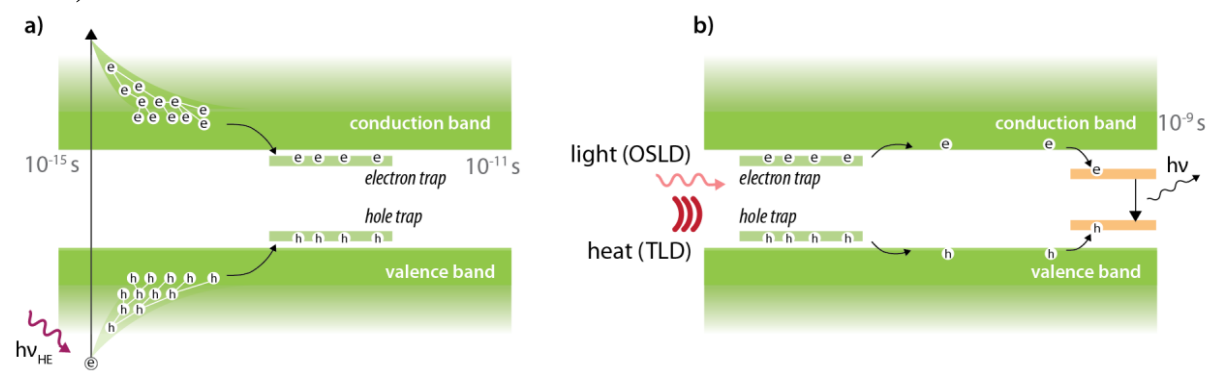

Figure 6. Luminescence in thermally and optically stimulated luminescent detectors. **a)** Irradiation leads to promotion of electrons from the valence band to conduction band. Electron and holes are subsequently trapped at trap centers which are introduced by doping impurities into crystals. **b)** An external stimulus in the form of heat or light is then provided to facilitate electron-hole recombination at luminescent center, which leads to production of optical photons.

When impurities are added to certain crystals, charge trapping occurs due to the added energy levels in the conduction-valence band gap. These additional energy levels act as traps for electrons and holes. Application of an external stimulus allows the trapped electrons and holes to escape allowing recombination at luminescent centers. It is this recombination process which leads to luminescence (Figure 6). Depending on the external stimulus, the dosimeters can be classified as TLD (thermo-luminescent dosimeter) or OSLD (optically stimulated luminescent dosimeter). The luminescence is considered to be delayed because the electrons and holes can remain trapped over long periods of time (sometimes up to thousands of years) and can only be read-out after stimulation. The practical implication of this is that real-time dosimetry is not feasible using TLD or OSLDs. In some cases however, certain materials such as europium-doped alkali halides, may exhibit short trap emptying (~25 ms) and luminescent decay times (~1 μs) which may render real-time dose monitoring possible[72].

Despite the lack of real-time readout, TLDs and OSLDs are of great importance in high dose-rate dosimetry because of their excellent dose-rate independence. In fact, one of the earliest studies of dose rate effect on skin toxicity in mice (1980) by Inada et al[73] used a lithium borate TLD to verify dose. They confirmed the lithium borate TLD to be independent up to a dose rate of 1.5 x $10^9$ Gy/s. Dose-rate dependency of TLDs has been investigated by multiple authors over the last few decades. Karzmarck et al[74] found LiF TLD to be dose-rate independent up to 2 x $10^6$ Gy/s. Tochilin and Goldstien [75] found the same TLD to be dose-rate independent up to 1.7 x $10^8$ Gy/s. More recently, Karsch et al[64] compared dose-rate independence of various detectors including TLDs and OSLDs. They found TLD and OSLD to be dose-rate independent up to 4 x $10^9$ Gy/s within 2%. In the context of FLASH, Jorge et al[34] compared LiF-100 TLD (Thermo Fisher, USA) against two dose-rate independent dosimeters, alanine and radiochromic film. The Oriatron eRT6 linac was used for this study with dose-rates ranging from 0.078 Gy/s up to 1500 Gy/s and the results are presented in Figure 7. Alanine, film, and the LiF-100 TLD were found to agree within 3% at all dose-rates.

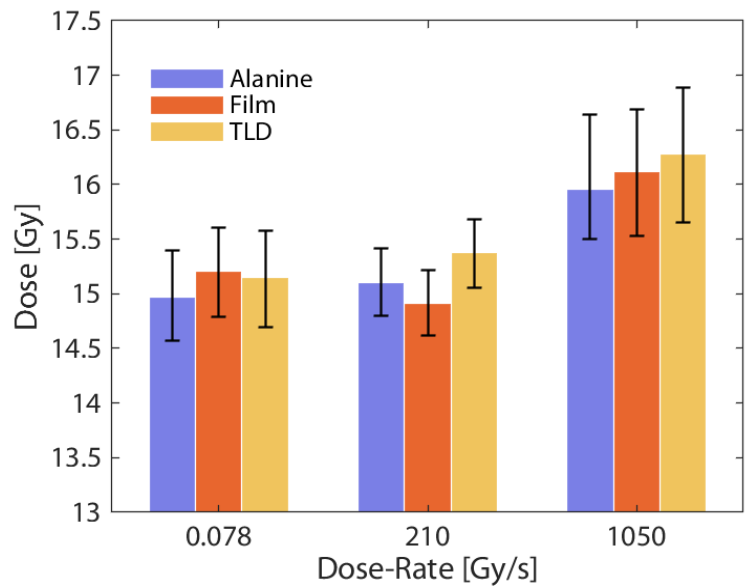

Figure 7. Measured absorbed dose under different dose-rate conditions for alanine, film and TLD. All detectors were found to agree within 3% at conventional and FLASH dose-rates, indicating excellent dose-rate independence[34].

In addition to dose-rate independence, TLD can be manufactured to be small and in powdered form. This is beneficial for small field dosimetry where high resolution is required. Additionally, the small form factor coupled with the ability to read out dose post irradiation renders TLDs as a viable tool for in-vivo dose verification. Indeed, for one FLASH study on mice whole brain irradiation[15], dosimetric verification was performed in-vivo using 3 x 3 x 1 mm³ TLD chips embedded inside the brain of a mouse cadaver at different points. A total dose of 10 Gy was either delivered in a single 1.8 us pulse or at conventional dose-rates (0.1Gy/s). The placement of the TLDs and the dose verification are shown in Figure 8. The dose measured by the TLDs agreed well with the prescribed dose of 10 Gy. Additionally, no dose rate effect was seen between measurements performed at 0.1 Gy/s compared to dose delivery in a single 1.8 us pulse (5 x 10⁷ G/s). This was one of the first non-superficial, in-vivo measurements performed at FLASH dose-rates.

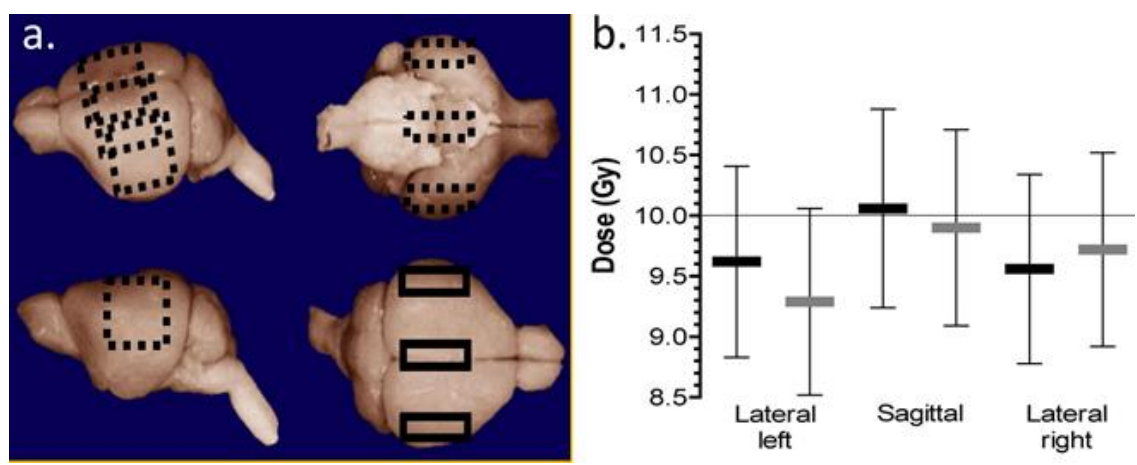

Figure 8. **a)** The three different positions where the TLDs were placed inside the brain of a mouse cadaver (sagittal/center, lateral left and right). **b)** The measured dose at different points. The black markers represent 10 Gy delivered in a single pulse of 1.8 us (FLASH-RT). The light gray markers represent the dose delivered at a dose-rate of 0.1Gy/s (i.e. conventional dose-rate). Error bars represent the relative uncertainty in the absorbed dose measurements (±8.2% in each case)[15]. Reprinted from "Irradiation in a flash: Unique sparing of memory in mice after whole brain irradiation with dose rates above 100Gy/s", Vol 124 / Issue 3, Figure 1, Copyright (2017), with permission from Elsevier.

One caveat of TLDs and OSLDs is that measurements are usually limited to a point. To overcome this, a few investigators have studied the possibility of using planar arrays of TLDs and OSLDs to measure spatial distribution of dose. One such TLD array was designed and tested at the European Synchrotron Radiation Facility (ESRF) by Ptaszkiewicz et al[76]. The study was primarily aimed at Microbeam Radiation Therapy (MRT); a novel external beam radiotherapy technique in which quasi-parallel beams with widths around 25-50 µm separated by 100-400 µm are delivered at ultra-high dose rates. The dose-rates in MRT can reach up to a few kGy/s[77]. The TLD array consisted of LiF:Mg,Cu,P 10 x 10 x 0.3 mm³ foils with different grain sizes (up to 150 µm). Dose read-out was performed using a 12-bit CCD camera with sub-millimeter resolution. Therefore, dose-rate independence and sub-millimeter resolution make this setup an attractive choice for FLASH-RT. Reusable 2D OSLD arrays have also been constructed with submillimeter resolution and large dynamic dose range[78,79].

Another passive luminescent detector of note is the Fluorescent Nuclear Track Detector (FNTD)[80]. FNTDs employ a single crystal of Aluminum Oxide doped with Magnesium and Carbon ($Al_2O_3$:C, Mg) with additional oxygen vacancy defects. The underlying physics is similar to OSLDs but with minor differences. Specifically, exposure to radiation produces new recombination or color centers which can then be probed non-destructively using microscopy techniques. In contrast, new recombination centers are not formed in OSLDs when exposed to radiation. This technique has been used for dosimetry in MRT[81] where spatial resolution of 1 μm was achieved. Additionally, FNTDs have also been tested to be dose-rate independent up to $10^8$ Gy/s and are capable of measuring dose over a large dynamic range ( 3 mGy to 100 Gy)[80]. Therefore, FNTDs are an attractive choice for dosimetry in FLASH-RT.

### 3.3.2) Scintillators

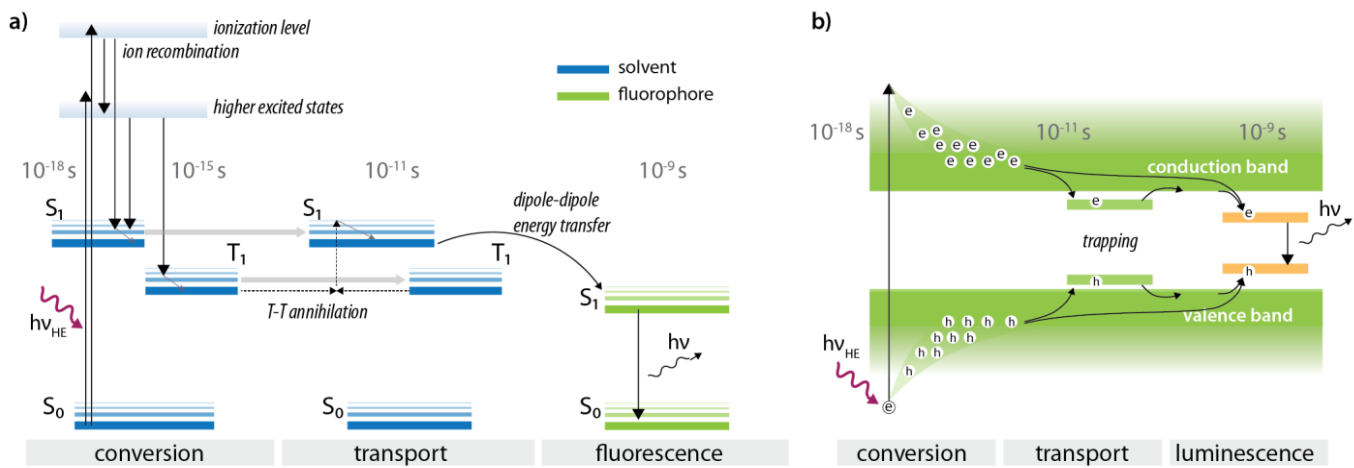

Figure 9. Typical processes of scintillation in **a)** organic and **b)** inorganic scintillators. The time scales over which these processes occur are also shown, indicating excellent temporal resolution. Figure b) was adapted from Nikl et al[82]

Scintillation is the phenomena by which an interaction of certain material (scintillator) with high energy photon or charged particle results in emission of optical photons. Scintillators can be broadly divided into two different categories 1) organic and 2) inorganic[83,84], with underlying physical mechanisms depicted in Figure 9. The process of scintillation in both material types follows a general mechanism composed of *conversion*, *transport (migration)*, and *luminescence*. Organic scintillators are typically aromatic hydrocarbon compounds, that produce excited states by ionizing radiation, and subsequently luminesce due to allowed π electron transitions between excited singlet state $S_{10}$ to various different vibrational sub-levels of the ground singlet state. When electronic transitions occur between singlet states, the emission and decay of luminescence is on the order of a few ns. However, most organic solutions do not exhibit high scintillation efficiency and are therefore used in conjunction with a solute. In this case, energy transfer occurs mainly via solvent-solvent interactions, and ultimately to the solute by dipole-dipole energy transfer[85]. In the case of polar solvent-based scintillators (e.g. quinine solution in water), ions and radicals are formed rather than excited states, and therefore ion recombination must occur prior solute excitation[86].

Inorganic scintillators typically consist of single or poly-crystalline materials, often doped with impurities that can act as luminescent centers. In the initial conversion phase, a large number of excited electrons and holes is created upon interaction of high energy photon or charged particle with the scintillator matrix, followed by thermalization and transport of created excited states to a luminescent center. Unlike in OSLD, an external stimulus to facilitate the release and recombination of electrons and holes is not needed due to the presence of an allowed transition at luminescent centers, and minimal number of charge traps. Similar to organic scintillators, rise and decay times for inorganic scintillators can also be on the order of a few ns.

Owing to their excellent tissue-equivalence and the ability to be miniaturized, multiple investigators have recommended the use of organic scintillators for small field dosimetry[21,39,87]. In particular, organic scintillators, such as the commercially available Exradin W1 (Standard Imaging) can be used as reference detectors for small fields against which correction factors for other detectors can be derived[39]. In contrast to organic type, inorganic scintillators are usually made with high-Z materials and are therefore not tissue-equivalent; a scenario which need to be accounted for in radiation dosimetry. Nonetheless, they have a role to play in FLASH-RT. Fast rise and decay times, radiation hardness and high detection efficiency due to increased photoelectric cross section for x-rays, makes inorganic scintillators an ideal tool for applications where superior time-resolution is required.

Typically, measurements performed with scintillators can be either point, planar 2D or 3D measurements. For point measurements, the setup usually consists of a small scintillator coupled to an optical fiber and a photodetector. Recently, Archer et al[88] demonstrated the use of miniature BG400 plastic scintillator (10 μm thick) coupled to a fiber optic and a SiPM for dosimetry at the Imaging and Medical Beam-Line at the Australian Synchrotron with an average dose-rate of 4435 Gy/s, resolving beams of 50 μm width.

In case of planar or 3D measurements, the setup typically consists of a scintillating volume imaged remotely at high spatial resolution (sub millimeter) with a CCD or a CMOS camera. The prompt emission of light, coupled with high frame-rate imaging capabilities of modern imaging sensors, make this technique suitable for online monitoring of machine output and dose delivery under FLASH irradiation conditions. Optical imaging of scintillation using cameras during external beam radiotherapy has already been widely implemented [84,89,90]. In a recent study[91], a time-gated intensified CMOS camera was used to image complex stereotactic radiosurgery (SRS) plans at high dose-rates in a radioluminescent phantom. The authors were able to resolve complex and highly modulated dose distributions spatially and temporally. Due to its high spatio-temporal resolution, optical imaging has also been used for quality assurance purposes in pencil beam scanning (PBS) proton therapy; a technique which also utilizes high dose rates (up to 200 Gy/s near the Bragg Peak). For example, Vigdor et al[92] used a xenon gas scintillator coupled to large PMTs for monitoring 2D beam characteristics in real-time for pulsed and pencil beam scanning proton radiotherapy treatments. The authors demonstrated a spatial resolution of a few hundred microns. Additionally, they noted that the gas scintillator was able to measure up to a dose rate of 350 Gy/s, whereas an ionization chamber started exhibiting ion-recombination effects at much smaller dose-rates. In another study, Darne et al[93,94] used three CMOS cameras to image proton pencil beam scanning inside a phantom filled with a liquid scintillator. The authors were able to image at 91 frames per second with sub millimeter resolution. More recently, Rahman et al[32] were able to resolve spatio-temporal (10 ms and 1 mm resolution) dose-rate dynamics up to 26 Gy/s for proton PBS using a scintillating sheet and a CMOS camera. As shown in Figure 10, the imaging technique visualized the proton beam parameters that modulated dose rate distributions and introduced cumulative dose rate histograms that can potentially be used for optimizing dose rate distributions for patient planning in FLASH-RT. One FLASH study by Favaudon et al[30], used a 2D scintillating array coupled to a CCD camera ( Lynx ® , IBA) for monitoring beam profiles. The scintillating screen was a 0.5 mm thick gadolinium based plastic material with an active area of 300 x 300 mm$^2$ and a spatial resolution of 0.5 mm. The detector was primarily used to assess field size, field homogeneity and dose linearity of the system and exhibited excellent linearity with increasing dose. The field homogeneity and FWHM values measured by the scintillating detector were within 1% at high and low dose rates. Peak dose-rate used in the study was around 2.4-3.5 x 10$^6$ Gy/s. The Lynx ® detector was also used by Beyreuther et al[13] for measuring field homogeneity in their proton FLASH setup. Multiple investigators have now used cameras and scintillation to reconstruct dose in 3D[95–98]. In the context of FLASH, this implies that a dose-rate independent detector that can measure dose in 3D with high spatial and temporal resolution ought to be able to measure dose-rate distribution in 3D in real-time. This information can potentially be used to predict the spatial distribution of the protective effect of FLASH in patient geometry.

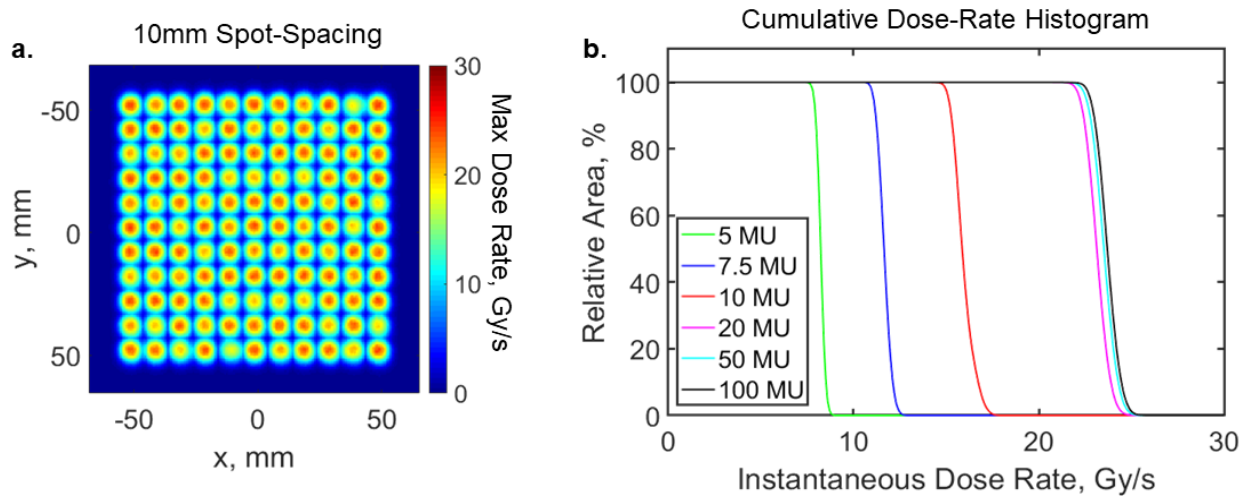

Figure 10. Scintillating sheet imaged dose rates from a proton pencil beam scanning system with modulated beam parameters in the treatment plans. **a.** Maximum dose rate distribution with 10 mm spot-spacing **b.** Cumulative dose rate histogram for varying minimum spot weight of a treated layer[32].

### 3.3.3) Cherenkov Radiation

**a)**

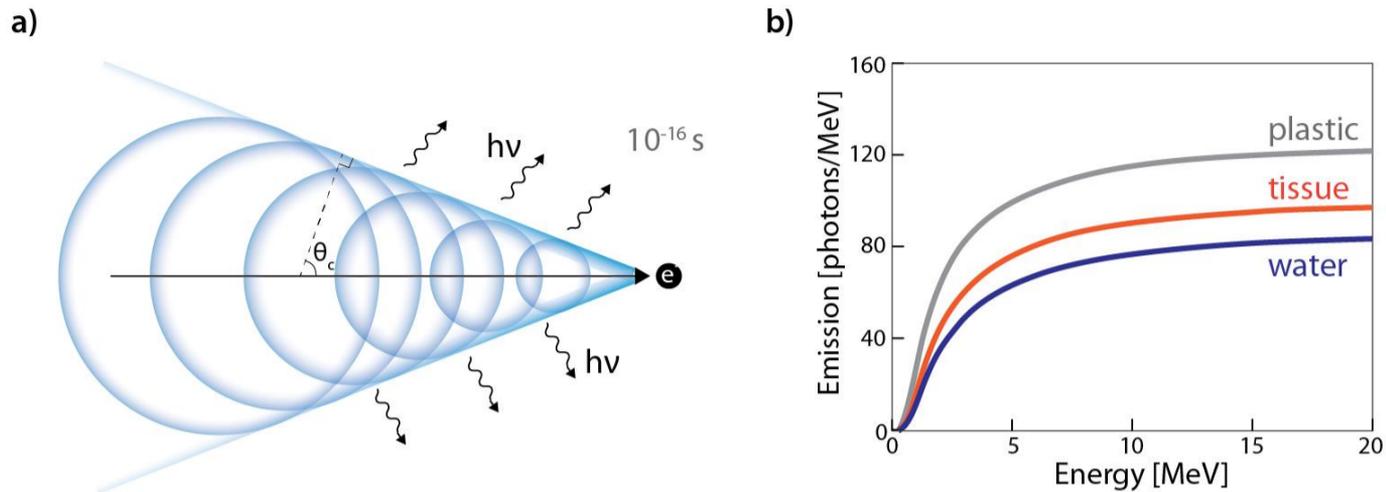

**b)**

Figure 11. **a)** Huygens representation of Cherenkov radiation mechanism in dielectric medium. Light is generated in a cone at caustic angle $\theta_c$ around the trajectory of charged particle. **b)** Energy dependence of Cherenkov radiation for different materials (adapted from Glaser et al[99])

Cherenkov radiation is the emission of optical photons in a dielectric medium when a charged particle travels at a phase velocity that exceeds the phase velocity of light in the medium. Electromagnetic fields associated with a charged particle can polarize the medium. If the particle moves slowly (relative to the speed of light in the medium), the relaxing dipoles experience a net destructive interference, and no light is emitted. However, if the particle's phase velocity exceeds that of light, asymmetric polarization can occur along the particle's trajectory, and the relaxing dipoles radiate energy with a net constructive interference, observed as visible Cherenkov radiation (Figure 11a). In contrast to scintillation, the Cherenkov light is not emitted isotropically, but rather in a cone with its axis aligned with particle trajectory. It has been shown by multiple groups that above the threshold for Cherenkov (261 keV in water) generation, the intensity of light emitted is proportional to dose[99–101], albeit with prominent energy dependence for particles below ~1 MeV. Importantly, Cherenkov light is created instantaneously[102] (~$10^{-12}$ s) upon interaction of the charged particle with the dielectric medium; this is faster than what most scintillators are capable of because of the various non-radiative mechanisms specific to the process of scintillation. Multiple investigators have now made use of Cherenkov radiation as time-of-flight PET detectors[102–104] due to its fast time response. Cherenkov radiation has also been found use in pulse radiolysis studies with pico[105] and femtosecond[106] time resolution. Additionally, Cherenkov radiation has been imaged in real-time[107(p),108] during multiple clinical radiotherapy treatments. The prompt nature of light emission, along with dose linearity makes Cherenkov emission an ideal tool for real-time dose monitoring. The general experimental setup for Cherenkov based dosimetric imaging is similar to the ones discussed earlier for scintillation dosimetry; an undoped optical fiber (i.e. production of Cherenkov and no scintillation) coupled to a photodetector or a volume capable of producing Cherenkov radiation imaged remotely with a camera. In the latter case, if a water phantom is subjected to radiation, Cherenkov emission can then be considered to be a water-equivalent dosimeter. However, due to inherent threshold below which no Cherenkov photons are generated, Cherenkov based detectors are expected to be energy dependent; a scenario which is not ideal for radiation dosimetry (Figure 11b).

In the context of FLASH, a Cherenkov probe was used for online monitoring of dose by Favaudon et al[30]. A number of tests were performed to confirm the efficacy of the Cherenkov detector. In one of the tests, a single 1 μs pulse of 3.9 or 5.0 MeV electrons were delivered to the probe. The area under the signal detected by the PMT (voltage against time) was found to be proportional to the energy of the beam. In another test, single pulses were delivered with increasing pulse widths (0.1 to 2.2 μs), which essentially translates into changing dose. The authors noted that the integral Cherenkov emission increased with beam energy, pulse duration and dose, without any saturation effects. Based on these results, the authors concluded that Cherenkov radiation has potential to be a useful tool for online-dose monitoring under high and low dose-rate conditions.

## 4. Biological Effects and Dosimetry: OER and LET

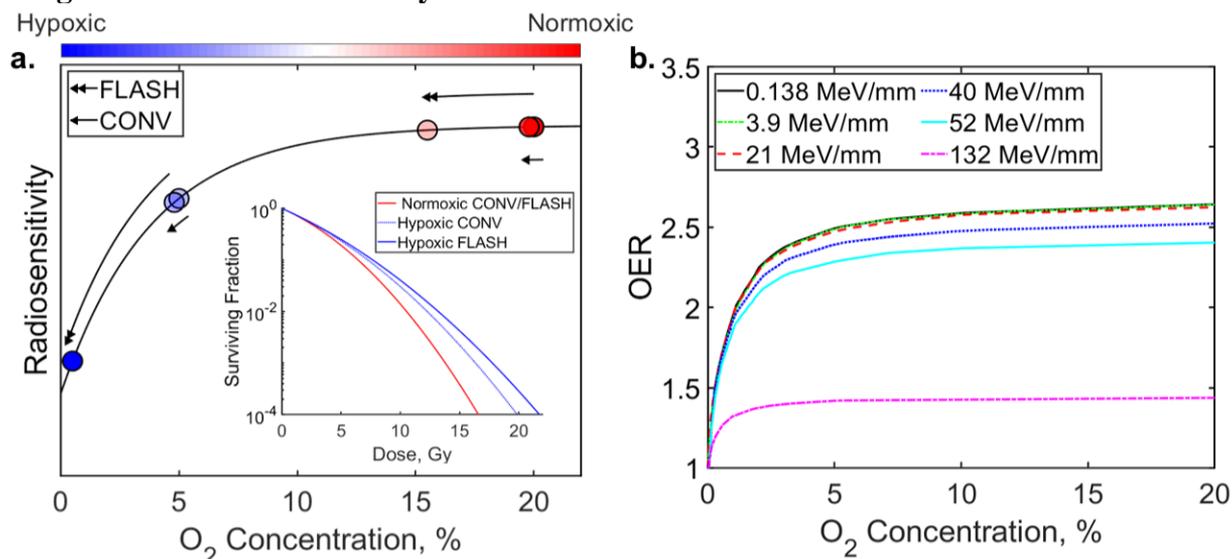

Figure 12 a) Radiosensitivity and cell surviving fraction dependency on oxygen concentration for FLASH and conventional (CONV) radiation therapy. Models are from Pratx et al and Adrien et al for normoxic (20% oxygen concentration) and hypoxic (5% oxygen concentration) cells. b) Oxygen enhancement ratio with $O_2$ concentration for irradiation with different LET [37,109–111]

A complete overview of the radiobiological underpinning of the FLASH effect is outside the scope of this study. However, the role of oxygen depletion will be briefly discussed here, since it is considered to be one of the major factors mediating the FLASH effect. Presence of molecular oxygen is known to make cells more susceptible to damage by radiation, as shown in the radiosensitivity curve in Figure 12a. This can be defined in terms of the oxygen enhancement ratio (OER), which is the ratio of dose needed to achieve the same biological effect under hypoxic and normoxic conditions. At ultra-high dose rates, it is hypothesized that transient hypoxia occurs which confers a protective effect on normal tissue. The improved differential response between tumor and normal tissue arises because the microenvironment surrounding solid tumors is already hypoxic[112] and remains largely unaffected by the depletion of oxygen. The time-scales over which oxygen depletion and reoxygenation occurs is important, since the underlying assumption is that oxygen is depleted at a rate faster than it can diffuse back into the normal tissue. Adrian et al[109] *in vitro* study supports the model and their results indicated there was no difference between cell death in hypoxic (5% oxygen concentration) cells, however cells under normoxic (20% oxygen concentration) oxygen conditions, showed increased survival from FLASH compared to conventional irradiation[109]. This can be attributed to larger gradient in radiosensitivity at normoxic oxygen concentration, thus a more prominent FLASH effect.

Luminescence imaging, in addition to dose and dose rate, can measure oxygen concentration and play an important role in testing the hypothesis *in vivo*. There are many indirect methods of estimating oxygen concentration in tissue including quantifying vascular parameters (intercapillary distance, distance from tumor cells to nearest vessel), perfusion, gene expression, protein levels, metabolism, DNA damage[113]. However, there are only a few direct methods of measuring oxygen concentration or tension directly, including the standard procedure of using a polarographic needle electrode system[114–117]. Collingridge et al[118] compared the standard polarographic method to an oxygen sensing system based on a time-resolved luminescence optical probe. The method relied on measuring the lifetime of the luminescence molecules from oxygen-quenching in the tip of the optical fiber and relating it to oxygen concentration. The authors confirmed that the time-resolved luminescence probe had the same degree of accuracy as a polarographic electrode system in measuring oxygen concentration. However, probes are invasive, measure at single points, and are scanned across the tissue to provide a histogram of oxygen concentration. Imaging techniques provide methods of quantifying oxygen concentration distributions. Positron emission tomography (PET) with $F^{18}$ labelled markers has been used to image hypoxia, but scans may take 2-4 hours, which is much longer than the time scale of FLASH effects[113]. Alternatively, $F^{19}$ based oximetry and magnetic resonance imaging can provide oxygen concentration at the multiple pixel level[119]. Electron paramagnetic resonance spectroscopy and imaging with oxygen sensitive particulate or water-soluble probes can also provide direct measurement of pO2 or [O2] with sub-millimeter spatial resolution and temporal resolution $<<1s$[120,121]. More recently, Cherenkov excited luminescent imaging (CELI) provided in vivo oxygen pressure maps based on lifetime imaging of fluorophore platinum(II)-G4 (PtG4)[122,123]. The Stern-Volmer equation was used to relate the decay of PtG4 to oxygen tension in tumors and normal tissue (pre and post euthanasia) (Figure 13). The imaging technique achieved submillimeter resolution of $pO_2$ across the surface and near sub-surface of tissue and can potentially image $pO_2$ post irradiation from a FLASH beam. This could be used to help relate the dose rate distribution to the oxygen depletion distribution in tissue and effects on the OER.

The ability of certain luminescent based detectors to quantify linear energy transfer (LET) can also play a crucial role for FLASH-RT. The OER is dependent on particle type and LET as shown in Figure 12b[110]. Heavy charged particles such as protons and alpha particles can have higher LET than photons or electrons and can reduce the effects of the OER. Thus, the oxygen depletion hypothesis brings into question whether heavy charged particles will provide the same degree of FLASH effects as electrons and photon FLASH beams. Furthermore, LET distribution of heavy charged particles are not homogeneous and have a drastic increase at the Bragg peak of the beam. So, quantifying spatial distribution of the LET may be important in describing the differences in FLASH effects of heavy particles and photons/electron beams. Currently, Monte Carlo methods or analytical methods are often used to determine LET distribution for treatment plans[124,125]. However, only certain detectors can *measure* LET of charged particle beams and majority of them are based on luminescent techniques. Fluorescence nuclear track detectors (FNTD) have been used to measure LET of individual proton tracks.[80,126]. However, FNTDs have a limited range of LET it can detect (5MeV/mm-1000MeV/mm), which does not include the range of LET distribution of proton beams. Alternatively, OSLD/TLD response dependency on LET can be utilized to determine both LET and dose distribution[127,128]. Nonetheless, these methods are passive and do not quantify dose or LET distribution in real time. Alsanea et al[129] showed that variable LET scintillation quenching in two different tissue equivalent organic scintillators can be utilized to determine dose and LET distribution in real time[129]. This method relies on Birk's law of scintillation quenching, requires a large difference in the quenching parameter, and requires the scintillators to be made of the same material (i.e. electron density). To note, silicon detectors have also been used to determine the mean LET of the proton and show its dependency on clinical proton beam energies (1-194MeV)[130].

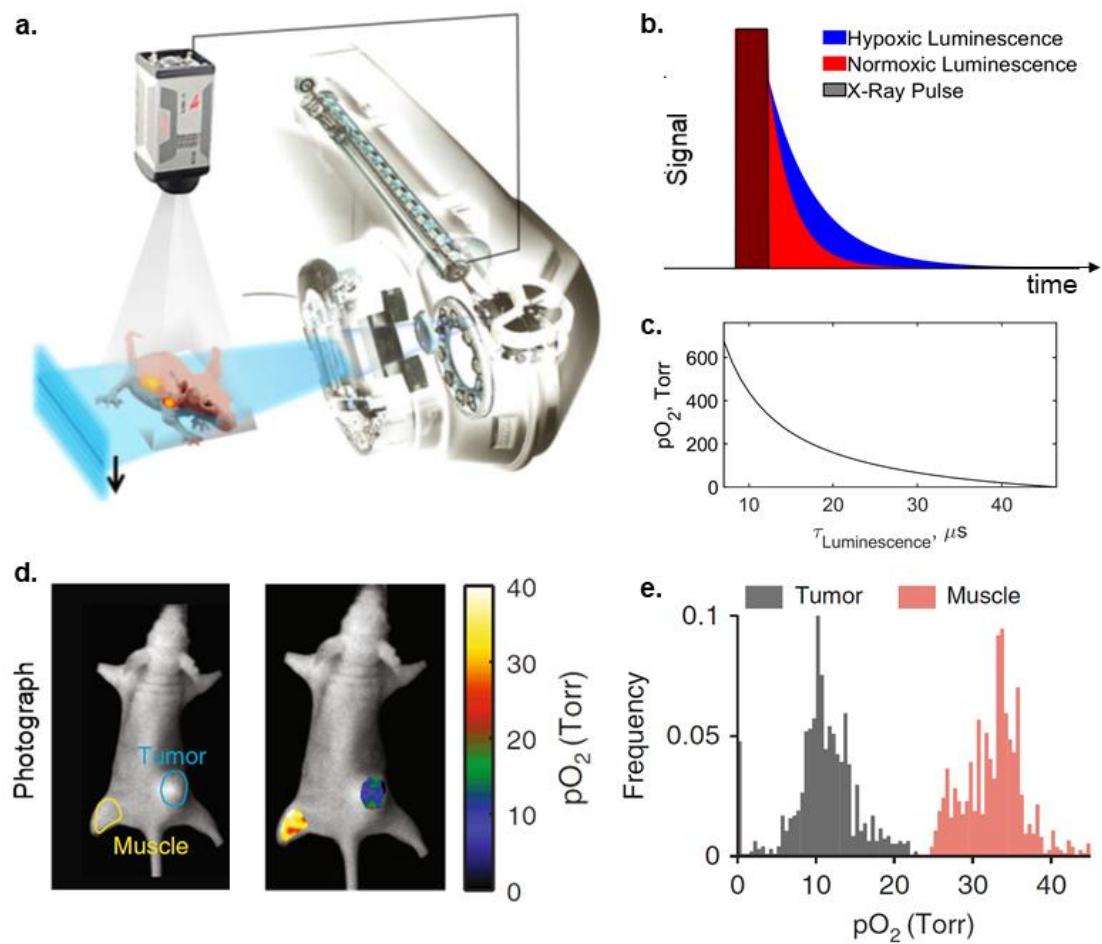

Figure 13. Cherenkov excited luminescence imaging of oxygen concentration; **a)** Schematic of a 6MV x-ray beam from the linear accelerator used to excite luminescence and captured by a time-synchronized camera; **b)** Time signatures of the x-ray pulse and the imaged luminescence in hypoxic and normoxic tissue; **c)** Relation between $pO_2$ pressure and luminescent lifetime based on Stern-Volmer model; **d)** $pO_2$ map of muscle and tumor tissue with their respective histograms shown in **e)**. [123,131,132]

## 4. Discussion and Conclusion

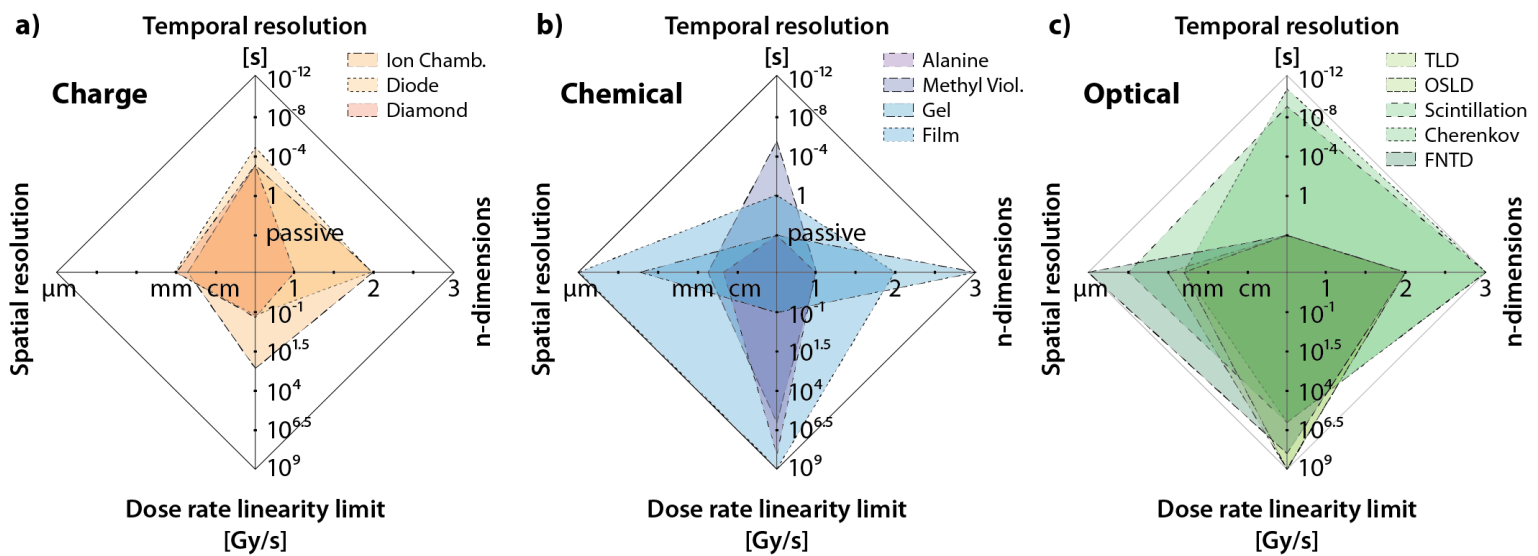

Figure 14. Dosimeter features relevant to FLASH-RT plotted as spider web plots for the three different categories of detectors/tools, including: **a)** charge-based dosimetry, **b)** chemical dosimetry/actinometry, and **c)** optical-based dosimetry.

A comprehensive list of dosimeters used in FLASH studies and other high dose-rate modalities is given in Table 1. The different columns represent some of the major issues identified in section 2. The values are based on typical values and usage encountered in literature. Exception to these values do exist; for example, radiochromic film is categorized as a passive detector, but attempts are real-time dosimetry with film has been made in the past[67,68]. The 'Measurement Type' column has bold entries in it, which indicate the way those dosimeters were employed in FLASH-RT studies. The time resolution values are based on the underlying physics of the dosimeters, as explained previously. This does not take in to account the available bandwidth of the read-out method. Of course, the dead-time of the read-out electronics should be considered while dealing with such dose-rates. While some of these issues are not necessarily unique to FLASH-RT, they nonetheless contribute to the overall dosimetric uncertainty.

Based on the unique temporal beam characteristics of FLASH-RT, dose-rate dependence, spatial resolution (in particular, accurate measurement of spatial distribution of dose-rate in a broad-beam), and time-resolution parameters shall be emphasized. To compare detectors based on these parameters (and the number of dimensions it has measured up to in literature), a spider chart in presented in Figure 14, where a), b) and c) refer to the three different categories of detectors. It can be seen that all luminescence based dosimeters exhibit excellent dose-rate independence. For chemical dosimeters, radiochromic film and alanine dosimeters also show dose-rate independence up to $10^9$ Gy/s. Even though methyl viologen is shown to be dose-rate independent up to a high dose-rate in Figure 14, they tend to be dose-rate dependent at really low dose-rates or low doses per pulse conditions because of the diffusion/decay of radiation induced species with time. Therefore, such chemical dosimeters, while promising at high dose-rates, might not be suitable if they are to quantify an in-homogeneous distribution of dose-rate. Charged based dosimeters tend to have a complex dependence on dose rate and are highly dependent on the temporal characteristics of the beam.

For measurement of dose in real-time, it can be seen that, luminescent detectors again tend to be superior when compared to chemical and charge based detectors. Most scintillator based detectors provide ~ns resolution, whereas Cherenkov radiation in this regard provides the best theoretical time-resolution (~ps); a fact which makes it an ideal candidate for online-monitoring of machine output without suffering from issues such as saturation or dose-rate dependence. Based on results presented by Favaudon et al[30], it can be argued that Cherenkov detectors can play a role similar to that of monitor chambers in conventional radiotherapy. Other luminescent detectors, such as TLD/OSLD and FNTD are suitable for passive measurements. Nonetheless, they still have a role to play in FLASH-RT, because of their dose-rate independence. Most charge-based dosimeters also offer decent temporal resolution; however, they are limited by their dependence on dose per pulse/dose-rate. Chemical based dosimeters tend to be feasible only for offline measurements, due to cumbersome read-out methods and the general instability (temporal and spatial) of radiation induced species.

Due to an increase in use of small radiation fields, detectors have been miniaturized to the extent, such that a resolution of ~1 mm is achievable with most modern detectors. It is important to distinguish between spatial resolution of point detectors from that of imaging detectors. For point detector, the spatial resolution is defined in terms of the spatial extent of the sensitive volume. For imaging detectors, the inter-detector spacing is perhaps a more suitable measure of spatial resolution. While, point solid state detectors, such as diamonds and diodes can indeed be constructed to be small, imaging arrays based on these detectors typically tend to exhibit an inter-detector spacing of 3-5 mm. Dose-rate dependence coupled with sparse detector spacing, makes these imaging arrays unsuitable for FLASH purposes. Additionally, solid-state devices tend to be non-tissue equivalent and energy dependent which can further complicate dosimetry due to small field issues. Comparing chemical and luminescent detectors, it can be seen that radiochromic film provide the best possible spatial resolution. High spatial resolution, tissue equivalence and dose-rate independence make radiochromic films an ideal tool for measuring spatial distribution of dose-rate in FLASH-RT. However, luminescent based detectors based on optical imaging techniques can provide the aforementioned qualities of radiochromic film, with the added advantage of high temporal resolution which makes real-time dose monitoring possible.

In addition to traditional dosimetry, the bio-chemical dose response of FLASH-RT was also discussed. In particular, it was shown that luminescent techniques can sense oxygen tension in real-time and can also measure dose and LET simultaneously for particle therapy. These parameters are crucial to understanding the underlying radiobiological mechanisms of the protective effect of FLASH. Questions such as how the FLASH effect varies with LET, oxygen concentration etc. can be answered using these techniques. In conclusion luminescence was presented as a tool which can play a diverse role in the performing dosimetry and understanding the FLASH effect caused by ultra-high dose-rates.

Table 1. Dosimeters and their capabilities rated for potential FLASH-RT dose measurement of key parameters.

| Response | Detectors | Instantaneous Dose-Rate/ Dose per Pulse ($D_p$) Dependence | Spatial Resolution | Time-Resolution | Energy Dependence | Measurement Type | FLASH Study |
|---|---|---|---|---|---|---|---|
| **Luminescence** | TLD/OSLD | Independent ($\sim 10^9$ Gy/s)[74,133] | ~ 1 mm | Passive | Tissue-equivalent | **1D**, 2D | e[15,34,65] |
| | Scintillators | Independent ($\sim 10^6$ Gy/s)[30] | ~ 1 mm | ~ns | Tissue-equivalent | 1D,**2D**,3D | p[13,18] |
| | Cherenkov | Independent ($\sim 10^6$ Gy/s)[30] | ~ 1 mm | ~ps | Energy Dependent | **1D**,2D,3D | e[30] |
| | FNTD | Independent ($\sim 10^8$ Gy/s)[80] | ~ 1 μm | Passive | Energy Dependent | 2D | NA |
| **Charge** | Ionization Chambers | Dependent on $D_p$[41,46] (>1 Gy/pulse) | ~3-5 mm | ~ms | Energy Dependence shows up > 2 MeV | **1D,2D** | p[13,18,19] e[15,34,65] ph[16,17] |
| | Diamonds | Dependent on $D_p$ (>1 mGy/pulse)[42] | ~ 1 mm | ~μs | Tissue-equivalent | **1D** | p[18] |
| | Si Diode | Dependent on $D_p$[48] (Independent ~0.2 Gy/s)[134] | ~ 1 mm | ~ms | Energy Dependent | **1D**,2D | NA |
| **Chemical** | Alanine Pellets | Independent ($10^8$ Gy/s)[63] | ~ 5 mm | Passive | Tissue-equivalent | **1D** | e[12,15,34,135] |
| | Methyl Viologen/ Fricke | Depends on the decay rate and diffusion of radiation induced species | ~ 2 mm | ~ns | Tissue-equivalent | **1D** | e[30,41] |
| | Radiochromic Film | Independent ($10^9$ Gy/s)[64,65] | ~1 μm | Passive | Tissue-equivalent | **2D** | p[18,19] e[10-12,15,31,34,65,136] ph[16] |
| | Gel Dosimeters | Strong dependence below 0.001 Gy/s[137] and above 0.10 Gy/s[138] | ~1 mm | Passive | Tissue-equivalent | 3D | NA |

**Acknowledgments**
This work was funded by National Institutes of Health grants R01 EB023909, R01 EB024498 and P01 CA023108.

**References:**

1. Hornsey S, Bewley DK. Hypoxia in Mouse Intestine Induced by Electron Irradiation at High Dose-rates. *International Journal of Radiation Biology and Related Studies in Physics, Chemistry and Medicine*. 1971;19(5):479-483. doi:10.1080/09553007114550611

2. Chang DS, Lasley FD, Das IJ, Mendonca MS, Dynlacht JR. Therapeutic Ratio. In: Chang DS, Lasley FD, Das IJ, Mendonca MS, Dynlacht JR, eds. *Basic Radiotherapy Physics and Biology*. Springer International Publishing; 2014:277-282. doi:10.1007/978-3-319-06841-1_27

3. Hall EJ, Brenner DJ. The dose-rate effect revisited: radiobiological considerations of importance in radiotherapy. *Int J Radiat Oncol Biol Phys*. 1991;21(6):1403-1414. doi:10.1016/0360-3016(91)90314-t

4. Town CD. Radiobiology. Effect of high dose rates on survival of mammalian cells. *Nature*. 1967;215(5103):847-848. doi:10.1038/215847a0

5. Prempree T, Michelsen A, Merz T. The repair time of chromosome breaks induced by pulsed x-rays on ultra-high dose-rate. *Int J Radiat Biol Relat Stud Phys Chem Med*. 1969;15(6):571-574. doi:10.1080/09553006914550871

6. Nias AH, Swallow AJ, Keene JP, Hodgson BW. Effects of pulses of radiation on the survival of mammalian cells. *Br J Radiol*. 1969;42(499):553. doi:10.1259/0007-1285-42-499-553-b

7. Berry RJ, Hall EJ, Forster DW, Storr TH, Goodman MJ. Survival of mammalian cells exposed to x rays at ultra-high dose-rates. *Br J Radiol*. 1969;42(494):102-107. doi:10.1259/0007-1285-42-494-102

8. Watts ME, Maughan RL, Michael BD. Fast kinetics of the oxygen effect in irradiated mammalian cells. *Int J Radiat Biol Relat Stud Phys Chem Med*. 1978;33(2):195-199. doi:10.1080/09553007814550091

9. Favaudon V, Caplier L, Monceau V, et al. Ultrahigh dose-rate FLASH irradiation increases the differential response between normal and tumor tissue in mice. *Sci Transl Med*. 2014;6(245):245ra93. doi:10.1126/scitranslmed.3008973

10. Levy K, Natarajan S, Wang J, et al. FLASH irradiation enhances the therapeutic index of abdominal radiotherapy in mice. *bioRxiv*. Published online December 12, 2019:2019.12.12.873414. doi:10.1101/2019.12.12.873414

11. Simmons DA, Lartey FM, Schüler E, et al. Reduced cognitive deficits after FLASH irradiation of whole mouse brain are associated with less hippocampal dendritic spine loss and neuroinflammation. *Radiotherapy and Oncology*. 2019;139:4-10. doi:10.1016/j.radonc.2019.06.006

12. Vozenin M-C, De Fornel P, Petersson K, et al. The Advantage of FLASH Radiotherapy Confirmed in Mini-pig and Cat-cancer Patients. *Clin Cancer Res*. 2019;25(1):35-42. doi:10.1158/1078-0432.CCR-17-3375

13. Beyreuther E, Brand M, Hans S, et al. Feasibility of proton FLASH effect tested by zebrafish embryo irradiation. *Radiotherapy and Oncology*. 2019;139:46-50. doi:10.1016/j.radonc.2019.06.024

14. Fouillade C, Curras-Alonso S, Giuranno L, et al. FLASH Irradiation Spares Lung Progenitor Cells and Limits the Incidence of Radio-induced Senescence. *Clin Cancer Res*. 2020;26(6):1497-1506. doi:10.1158/1078-0432.CCR-19-1440

15. Montay-Gruel P, Petersson K, Jaccard M, et al. Irradiation in a flash: Unique sparing of memory in mice after whole brain irradiation with dose rates above 100 Gy/s. *Radiotherapy and Oncology*. 2017;124(3):365-369. doi:10.1016/j.radonc.2017.05.003

16. Montay-Gruel P, Bouchet A, Jaccard M, et al. X-rays can trigger the FLASH effect: Ultra-high dose-rate synchrotron light source prevents normal brain injury after whole brain irradiation in mice. *Radiother Oncol*. 2018;129(3):582-588. doi:10.1016/j.radonc.2018.08.016

17. Smyth LML, Donoghue JF, Ventura JA, et al. Comparative toxicity of synchrotron and conventional radiation therapy based on total and partial body irradiation in a murine model. *Sci Rep*. 2018;8(1):12044. doi:10.1038/s41598-018-30543-1

18. Patriarca A, Fouillade C, Auger M, et al. Experimental Set-up for FLASH Proton Irradiation of Small Animals Using a Clinical System. *International Journal of Radiation Oncology*Biology*Physics*. 2018;102(3):619-626. doi:10.1016/j.ijrobp.2018.06.403

19. Diffenderfer ES, Verginadis II, Kim MM, et al. Design, Implementation, and in Vivo Validation of a Novel Proton FLASH Radiation Therapy System. *International Journal of Radiation Oncology*Biology*Physics*. 2020;106(2):440-448. doi:10.1016/j.ijrobp.2019.10.049

20. Das IJ, Ding GX, Ahnesjö A. Small fields: Nonequilibrium radiation dosimetry: Small fields: Nonequilibrium radiation dosimetry. *Medical Physics*. 2007;35(1):206-215. doi:10.1118/1.2815356

21. Parwaie W, Refahi S, Ardekani MA, Farhood B. Different Dosimeters/Detectors Used in Small-Field Dosimetry: Pros and Cons. *J Med Signals Sens*. 2018;8(3):195-203. doi:10.4103/jmss.JMSS_3_18

22. Grotzer MA, Schültke E, Bräuer-Krisch E, Laissue JA. Microbeam radiation therapy: Clinical perspectives. *Physica Medica*. 2015;31(6):564-567. doi:10.1016/j.ejmp.2015.02.011

23. Bräuer-Krisch E, Adam J-F, Alagoz E, et al. Medical physics aspects of the synchrotron radiation therapies: Microbeam radiation therapy (MRT) and synchrotron stereotactic radiotherapy (SSRT). *Physica Medica*. 2015;31(6):568-583. doi:10.1016/j.ejmp.2015.04.016

24. Draeger E, Sawant A, Johnstone C, et al. A Dose of Reality: How 20 Years of Incomplete Physics and Dosimetry Reporting in Radiobiology Studies May Have Contributed to the Reproducibility Crisis. *International Journal of Radiation Oncology • Biology • Physics*. 2020;106(2):243-252. doi:10.1016/j.ijrobp.2019.06.2545

25. Pedersen KH, Kunugi KA, Hammer CG, Culberson WS, DeWerd LA. Radiation Biology Irradiator Dose Verification Survey. *rare*. 2016;185(2):163-168. doi:10.1667/RR14155.1

26. Desrosiers M, DeWerd L, Deye J, et al. The Importance of Dosimetry Standardization in Radiobiology. *J RES NATL INST STAN*. 2013;118:403. doi:10.6028/jres.118.021

27. Wilson JD, Hammond EM, Higgins GS, Petersson K. Ultra-High Dose Rate (FLASH) Radiotherapy: Silver Bullet or Fool's Gold? *Front Oncol*. 2020;9. doi:10.3389/fonc.2019.01563

28. Bourhis J, Montay-Gruel P, Gonçalves Jorge P, et al. Clinical translation of FLASH radiotherapy: Why and how? *Radiother Oncol*. 2019;139:11-17. doi:10.1016/j.radonc.2019.04.008

29. Marlen P van, Dahele M, Folkerts M, Abel E, Slotman BJ, Verbakel WFAR. Bringing FLASH to the Clinic: Treatment Planning Considerations for Ultrahigh Dose-Rate Proton Beams. *International Journal of Radiation Oncology • Biology • Physics*. 2020;106(3):621-629. doi:10.1016/j.ijrobp.2019.11.011

30. Favaudon V, Lentz J-M, Heinrich S, et al. Time-resolved dosimetry of pulsed electron beams in very high dose-rate, FLASH irradiation for radiotherapy preclinical studies. *Nuclear Instruments and Methods in Physics Research Section A: Accelerators, Spectrometers, Detectors and Associated Equipment*. 2019;944:162537. doi:10.1016/j.nima.2019.162537

31. Schüler E, Trovati S, King G, et al. Experimental Platform for Ultra-high Dose Rate FLASH Irradiation of Small Animals Using a Clinical Linear Accelerator. *Int J Radiat Oncol Biol Phys*. 2017;97(1):195-203. doi:10.1016/j.ijrobp.2016.09.018

32. Rahman M, Bruza P, Langen KM, et al. Characterization of a new scintillation imaging system for proton pencil beam dose rate measurements. *Phys Med Biol*. Published online 2020. doi:10.1088/1361-6560/ab9452

33. van de Water S, Safai S, Schippers JM, Weber DC, Lomax AJ. Towards FLASH proton therapy: the impact of treatment planning and machine characteristics on achievable dose rates. *Acta Oncol*. 2019;58(10):1463-1469. doi:10.1080/0284186X.2019.1627416

34. Jorge PG, Jaccard M, Petersson K, et al. Dosimetric and preparation procedures for irradiating biological models with pulsed electron beam at ultra-high dose-rate. *Radiotherapy and Oncology*. 2019;139:34-39. doi:10.1016/j.radonc.2019.05.004

35. Petersson K, Adrian G, Butterworth K, McMahon SJ. A quantitative analysis of the role of oxygen tension in FLASH radiotherapy. *International Journal of Radiation Oncology*Biology*Physics*. Published online March 5, 2020. doi:10.1016/j.ijrobp.2020.02.634

36. Weiss H, Epp ER, Heslin JM, Ling CC, Santomasso A. Oxygen Depletion in Cells Irradiated at Ultra-high Dose-rates and at Conventional Dose-rates. *International Journal of Radiation Biology and Related Studies in Physics, Chemistry and Medicine*. 1974;26(1):17-29. doi:10.1080/09553007414550901

37. Pratx G, Kapp DS. A computational model of radiolytic oxygen depletion during FLASH irradiation and its effect on the oxygen enhancement ratio. *Phys Med Biol*. 2019;64(18):185005. doi:10.1088/1361-6560/ab3769

38.    Wuerfel JU. DOSE MEASUREMENTS IN SMALL FIELDS. Published online 2013:10.

39.    Casar B, Gershkevitsh E, Mendez I, Jurković S, Huq MS. A novel method for the determination of field output factors and output correction factors for small static fields for six diodes and a microdiamond detector in megavoltage photon beams. *Medical Physics*. 2019;46(2):944-963. doi:10.1002/mp.13318

40.    Palmans H, Andreo P, Huq MS, Seuntjens J, Christaki KE, Meghzifene A. Dosimetry of small static fields used in external photon beam radiotherapy: Summary of TRS-483, the IAEA–AAPM international Code of Practice for reference and relative dose determination. *Medical Physics*. 2018;45(11):e1123-e1145. doi:10.1002/mp.13208

41.    Petersson K, Jaccard M, Germond J-F, et al. High dose-per-pulse electron beam dosimetry - A model to correct for the ion recombination in the Advanced Markus ionization chamber. *Medical Physics*. 2017;44(3):1157-1167. doi:10.1002/mp.12111

42.    Brualla-González L, Gómez F, Pombar M, Pardo-Montero J. Dose rate dependence of the PTW 60019 microDiamond detector in high dose-per-pulse pulsed beams. *Phys Med Biol*. 2016;61(1):N11-N19. doi:10.1088/0031-9155/61/1/N11

43.    Saini AS, Zhu TC. Dose rate and SDD dependence of commercially available diode detectors. *Medical Physics*. 2004;31(4):914-924. doi:10.1118/1.1650563

44.    Boag JW. The recombination correction for an ionisation chamber exposed to pulsed radiation in a \textquotesingleswept beam\textquotesingle technique. I. Theory. *Phys Med Biol*. 1982;27(2):201–211. doi:10.1088/0031-9155/27/2/001

45.    Fowler JF. Radiation-induced Conductivity in the Solid State, and Some Applications. *Phys Med Biol*. 1959;3(4):395-410. doi:10.1088/0031-9155/3/4/307

46.    Ade N, Nam TL, Derry TE, Mhlanga SH. The dose rate dependence of synthetic diamond detectors in the relative dosimetry of high-energy electron therapy beams. *Radiation Physics and Chemistry*. 2014;98:155-162. doi:10.1016/j.radphyschem.2014.02.003

47.    Fidanzio A, Azario L, Miceli R, Russo A, Piermattei A. PTW-diamond detector: dose rate and particle type dependence. *Med Phys*. 2000;27(11):2589-2593. doi:10.1118/1.1318218

48.    Jursinic PA. Dependence of diode sensitivity on the pulse rate of delivered radiation. *Medical Physics*. 2013;40(2):021720. doi:10.1118/1.4788763

49.    Neira S, Brualla-Gónzalez L, Prieto-Pena J, Gómez F, Pardo-Montero J. A kinetic model of diode detector response to pulsed radiation beams. *Phys Med Biol*. 2019;64(20):205007. doi:10.1088/1361-6560/ab4460

50.    Shi J, Simon WE, Zhu TC. Modeling the instantaneous dose rate dependence of radiation diode detectors. *Medical Physics*. 2003;30(9):2509-2519. doi:10.1118/1.1602171

51.    Müller O, Stötzel J, Lützenkirchen-Hecht D, Frahm R. Gridded Ionization Chambers for Time Resolved X-Ray Absorption Spectroscopy. *Journal of Physics: Conference Series*. 2013;425(9):092010. doi:10.1088/1742-6596/425/9/092010

52.    Schroder DK. Carrier lifetimes in silicon. *IEEE Transactions on Electron Devices*. 1997;44(1):160-170. doi:10.1109/16.554806

53.    Kozlov SF, Stuck R, Hage-Ali M, Siffert P. Preparation and Characteristics of Natural Diamond Nuclear Radiation Detectors. *IEEE Transactions on Nuclear Science*. 1975;22(1):160-170. doi:10.1109/TNS.1975.4327634

54.    Isberg J, Hammersberg J, Johansson E, et al. High carrier mobility in single-crystal plasma-deposited diamond. *Science*. 2002;297(5587):1670-1672. doi:10.1126/science.1074374

55.    Hendry JH, Moore JV, Hodgson BW, Keene JP. The constant low oxygen concentration in all the target cells for mouse tail radionecrosis. *Radiat Res*. 1982;92(1):172-181.

56.    Hu J, Rovey JL. Faraday cup with nanosecond response and adjustable impedance for fast electron beam characterization. *Rev Sci Instrum*. 2011;82(7):073504. doi:10.1063/1.3610649

57.    Richter C, Karsch L, Dammene Y, et al. A dosimetric system for quantitative cell irradiation experiments with laser-accelerated protons. *Phys Med Biol*. 2011;56(6):1529–1543. doi:10.1088/0031-9155/56/6/002

58. Prokůpek J, Kaufman J, Margarone D, et al. Development and first experimental tests of Faraday cup array. *Review of Scientific Instruments*. 2014;85(1):013302. doi:10.1063/1.4859496

59. Lerch MLF, Petasecca M, Cullen A, et al. Dosimetry of intensive synchrotron microbeams. *Radiation Measurements*. 2011;46(12):1560-1565. doi:10.1016/j.radmeas.2011.08.009

60. O'Leary M, Boscolo D, Breslin N, et al. Observation of dose-rate dependence in a Fricke dosimeter irradiated at low dose rates with monoenergetic X-rays. *Sci Rep*. 2018;8(1):1-9. doi:10.1038/s41598-018-21813-z

61. Anton M. Uncertainties in alanine/ESR dosimetry at the Physikalisch-Technische Bundesanstalt. *Phys Med Biol*. 2006;51(21):5419–5440. doi:10.1088/0031-9155/51/21/003

62. Kudoh H, Celina M, Kaye RJ, Gillen KT, Clough RL. Response of alanine dosimeters at very high dose rate. *Applied Radiation and Isotopes*. 1997;48(4):497-499. doi:10.1016/S0969-8043(96)00281-3

63. Soliman YS, Pellicioli P, Beshir WB, et al. A comparative dosimetry study of an alanine dosimeter with a PTW PinPoint chamber at ultra-high dose rates of synchrotron radiation. *Physica Medica: European Journal of Medical Physics*. 2020;71:161-167. doi:10.1016/j.ejmp.2020.03.007

64. Karsch L, Beyreuther E, Burris-Mog T, et al. Dose rate dependence for different dosimeters and detectors: TLD, OSL, EBT films, and diamond detectors. *Medical Physics*. 2012;39(5):2447-2455. doi:10.1118/1.3700400

65. Jaccard M, Petersson K, Buchillier T, et al. High dose-per-pulse electron beam dosimetry: Usability and dose-rate independence of EBT3 Gafchromic films. *Medical Physics*. 2017;44(2):725-735. doi:10.1002/mp.12066

66. McLaughlin WL, Al-Sheikhly M, Lewis DF, Kovács A, Wojnárovits L. Radiochromic Solid-State Polymerization Reaction. In: *Irradiation of Polymers*. Vol 620. ACS Symposium Series. American Chemical Society; 1996:152-166. doi:10.1021/bk-1996-0620.ch011

67. Casolaro P, Campajola L, Breglio G, et al. Real-time dosimetry with radiochromic films. *Scientific Reports*. 2019;9(1). doi:10.1038/s41598-019-41705-0

68. Rink A, Vitkin IA, Jaffray DA. Suitability of radiochromic medium for real-time optical measurements of ionizing radiation dose. *Medical Physics*. 2005;32(4):1140-1155. doi:10.1118/1.1877832

69. Baldock C, De Deene Y, Doran S, et al. Polymer gel dosimetry. *Phys Med Biol*. 2010;55(5):R1-R63. doi:10.1088/0031-9155/55/5/R01

70. Deene YD. Essential characteristics of polymer gel dosimeters. *J Phys: Conf Ser*. 2004;3:34-57. doi:10.1088/1742-6596/3/1/006

71. Khan M, Heilemann G, Kuess P, Georg D, Berg A. The impact of the oxygen scavenger on the dose-rate dependence and dose sensitivity of MAGIC type polymer gels. *Phys Med Biol*. 2018;63(6):06NT01. doi:10.1088/1361-6560/aab00b

72. McKeever SWS. Optically stimulated luminescence: A brief overview. *Radiation Measurements*. 2011;46(12):1336-1341. doi:10.1016/j.radmeas.2011.02.016

73. Inada T, Nishio H, Amino S, Abe K, Saito K. High dose-rate dependence of early skin reaction in mouse. *Int J Radiat Biol Relat Stud Phys Chem Med*. 1980;38(2):139-145. doi:10.1080/09553008014551031

74. Karzmark CJ, White J, Fowler JF. Lithium Fluoride Thermoluminescence Dosimetry. *Phys Med Biol*. 1964;9(3):273–286. doi:10.1088/0031-9155/9/3/302

75. Tochilin E, Goldstein N. Dose Rate and Spectral Measurements from Pulsed X-ray Generators. *Health Physics*. 1966;12(12):1705–1714.

76. Ptaszkiewicz M, Braurer-Kirsch E, Klosowski M, Czopyk L, Olko P. TLD dosimetry for microbeam radiation therapy at the European Synchrotron Radiation Facility. *Radiation Measurements*. 2008;43(2):990-993. doi:10.1016/j.radmeas.2007.12.050

77. Eling L, Bouchet A, Nemoz C, et al. Ultra high dose rate Synchrotron Microbeam Radiation Therapy. Preclinical evidence in view of a clinical transfer. *Radiotherapy and Oncology*. 2019;139:56-61. doi:10.1016/j.radonc.2019.06.030

78. Ahmed MF, Shrestha N, Ahmad S, Schnell E, Akselrod MS, Yukihara EG. Demonstration of 2D dosimetry using Al2O3 optically stimulated luminescence films for therapeutic megavoltage x-ray and ion beams. *Radiation Measurements*. 2017;106:315-320. doi:10.1016/j.radmeas.2017.04.010

79. Wouter C, Dirk V, Paul L, Tom D. A reusable OSL-film for 2D radiotherapy dosimetry. *Phys Med Biol*. 2017;62(21):8441-8454. doi:10.1088/1361-6560/aa8de6

80. Akselrod MS, Sykora GJ. Fluorescent nuclear track detector technology – A new way to do passive solid state dosimetry. *Radiation Measurements*. 2011;46(12):1671-1679. doi:10.1016/j.radmeas.2011.06.018

81. Bartz JA, Sykora GJ, Bräuer-Krisch E, Akselrod MS. Imaging and dosimetry of synchrotron microbeam with aluminum oxide fluorescent detectors. *Radiation Measurements*. 2011;46(12):1936-1939. doi:10.1016/j.radmeas.2011.04.003

82. Nikl M. Scintillation detectors for x-rays. *Meas Sci Technol*. 2006;17(4):R37-R54. doi:10.1088/0957-0233/17/4/R01

83. Birks JB. The Theory and Practice of Scintillation Counting. :671.

84. Beddar S, Beaulieu L. Scintillation Dosimetry. :424.

85. Horrocks DL. APPLICATIONS OF LIQUID SCINTILLATION COUNTING. :357.

86. Peng C-T, Horrocks DL, Alpen EL. *Liquid Scintillation Counting: Recent Applications and Development*. Academic Press; 1980.

87. Benmakhlouf H, Andreo P. Spectral distribution of particle fluence in small field detectors and its implication on small field dosimetry. *Medical Physics*. 2017;44(2):713-724. doi:10.1002/mp.12042

88. Archer J, Li E, Davis J, Cameron M, Rosenfeld A, Lerch M. High spatial resolution scintillator dosimetry of synchrotron microbeams. *Sci Rep*. 2019;9(1):1-7. doi:10.1038/s41598-019-43349-6

89. Beddar S, Archambault L, Sahoo N, et al. Exploration of the potential of liquid scintillators for real-time 3D dosimetry of intensity modulated proton beams. *Medical Physics*. 2009;36(5):1736-1743. doi:10.1118/1.3117583

90. Pönisch F, Archambault L, Briere TM, et al. Liquid scintillator for 2D dosimetry for high-energy photon beams: Liquid scintillator for 2D dosimetry. *Medical Physics*. 2009;36(5):1478-1485. doi:10.1118/1.3106390

91. Ashraf MR, Bruza P, Pogue BW, et al. Optical imaging provides rapid verification of static small beams, radiosurgery, and VMAT plans with millimeter resolution. *Medical Physics*. 0(0). doi:10.1002/mp.13797

92. Vigdor SE, Klyachko AV, Solberg KA, Pankuch M. A gas scintillator detector for 2D dose profile monitoring in pencil beam scanning and pulsed beam proton radiotherapy treatments. *Phys Med Biol*. 2017;62(12):4946-4969. doi:10.1088/1361-6560/aa6ce2

93. Darne CD, Alsanea F, Robertson DG, Hojo Y, Sahoo N, Beddar S. 3D Scintillator Detector System for Proton Scanning Beam Therapy. *International Journal of Radiation Oncology • Biology • Physics*. 2019;105(1):S89-S90. doi:10.1016/j.ijrobp.2019.06.563

94. Darne CD, Alsanea F, Robertson DG, Sahoo N, Beddar S. Performance characterization of a 3D liquid scintillation detector for discrete spot scanning proton beam systems. *Phys Med Biol*. 2017;62(14):5652-5667. doi:10.1088/1361-6560/aa780b

95. Goulet M, Rilling M, Gingras L, Beddar S, Beaulieu L, Archambault L. Novel, full 3D scintillation dosimetry using a static plenoptic camera. *Med Phys*. 2014;41(8). doi:10.1118/1.4884036

96. Kirov AS, Piao JZ, Mathur NK, et al. The three-dimensional scintillation dosimetry method: test for a 106Ru eye plaque applicator. *Phys Med Biol*. 2005;50(13):3063-3081. doi:10.1088/0031-9155/50/13/007

97. Beddar S. Real-time volumetric scintillation dosimetry. *J Phys: Conf Ser*. 2015;573:012005. doi:10.1088/1742-6596/573/1/012005

98. Bruza P, Andreozzi JM, Gladstone DJ, Jarvis LA, Rottmann J, Pogue BW. Online Combination of EPID Cherenkov Imaging for 3-D Dosimetry in a Liquid Phantom. *IEEE Transactions on Medical Imaging*. 2017;36(10):2099-2103. doi:10.1109/TMI.2017.2717800

99. Glaser AK, Zhang R, Gladstone DJ, Pogue BW. Optical dosimetry of radiotherapy beams using Cherenkov radiation: the relationship between light emission and dose. *Phys Med Biol*. 2014;59(14):3789-3811. doi:10.1088/0031-9155/59/14/3789

100. Glaser AK, Zhang R, Gladstone DJ, Pogue BW. Optical dosimetry of radiotherapy beams using Cherenkov radiation: the relationship between light emission and dose. *Phys Med Biol*. 2014;59(14):3789–3811. doi:10.1088/0031-9155/59/14/3789

101. Glaser AK, Davis SC, McClatchy DM, Zhang R, Pogue BW, Gladstone DJ. Projection imaging of photon beams by the Čerenkov effect: Projection imaging of photon beams by the Čerenkov effect. *Medical Physics*. 2012;40(1):012101. doi:10.1118/1.4770286

102. Brunner SE, Schaart DR. BGO as a hybrid scintillator / Cherenkov radiator for cost-effective time-of-flight PET. *Phys Med Biol*. 2017;62(11):4421-4439. doi:10.1088/1361-6560/aa6a49

103. Brunner SE, Gruber L, Marton J, Suzuki K, Hirtl A. Studies on the Cherenkov Effect for Improved Time Resolution of TOF-PET. *IEEE Transactions on Nuclear Science*. 2014;61(1):443-447. doi:10.1109/TNS.2013.2281667

104. Cates JW, Levin CS. Evaluation of a clinical TOF-PET detector design that achieves $\leqslant 100$ ps coincidence time resolution. *Phys Med Biol*. 2018;63(11):115011. doi:10.1088/1361-6560/aac504

105. Grigoryants VM, Lozovoy VV, Chernousov YuD, et al. Pulse radiolysis system with picosecond time resolution referred to Cherenkov radiation. *International Journal of Radiation Applications and Instrumentation Part C Radiation Physics and Chemistry*. 1989;34(3):349-352. doi:10.1016/1359-0197(89)90243-9

106. Yang J, Kondoh T, Yoshida A, Yoshida Y. Double-decker femtosecond electron beam accelerator for pulse radiolysis. *Review of Scientific Instruments*. 2006;77(4):043302. doi:10.1063/1.2195090

107. Glaser AK, Andreozzi JM, Davis SC, et al. Video-rate optical dosimetry and dynamic visualization of IMRT and VMAT treatment plans in water using Cherenkov radiation. *Med Phys*. 2014;41(6):062102. doi:10.1118/1.4875704

108. Jarvis LA, Zhang R, Gladstone DJ, et al. Cherenkov video imaging allows for the first visualization of radiation therapy in real time. *Int J Radiat Oncol Biol Phys*. 2014;89(3):615-622. doi:10.1016/j.ijrobp.2014.01.046

109. Adrian G, Konradsson E, Lempart M, Bäck S, Ceberg C, Petersson K. The FLASH effect depends on oxygen concentration. *Br J Radiol*. 2020;93(1106):20190702. doi:10.1259/bjr.20190702

110. Richardson RB. Age-dependent changes in oxygen tension, radiation dose and sensitivity within normal and diseased coronary arteries–Part C: Oxygen effect and its implications on high- and low-LET dose. *International Journal of Radiation Biology*. 2008;84(10):858-865. doi:10.1080/09553000802389686

111. Boscolo D, Krämer M, Fuss MC, Durante M, Scifoni E. Impact of Target Oxygenation on the Chemical Track Evolution of Ion and Electron Radiation. *International Journal of Molecular Sciences*. 2020;21(2):424. doi:10.3390/ijms21020424

112. Hockel M, Vaupel P. Tumor Hypoxia: Definitions and Current Clinical, Biologic, and Molecular Aspects. *JNCI Journal of the National Cancer Institute*. 2001;93(4):266-276. doi:10.1093/jnci/93.4.266

113. Horsman MR, Mortensen LS, Petersen JB, Busk M, Overgaard J. Imaging hypoxia to improve radiotherapy outcome. *Nature Reviews Clinical Oncology*. 2012;9(12):674-687. doi:10.1038/nrclinonc.2012.171

114. Vaupel P, Kelleher DK, H[ouml ]ckel M. Oxygenation status of malignant tumors: Pathogenesis of hypoxia and significance for tumor therapy. *Seminars in Oncology*. 2001;28(2F):29-35. doi:10.1053/sonc.2001.25398

115. Sørensen M, Horsman MR, Cumming P, Munk OL, Keiding S. Effect of intratumoral heterogeneity in oxygenation status on FMISO PET, autoradiography, and electrode Po2 measurements in murine tumors. *International Journal of Radiation Oncology*Biology*Physics*. 2005;62(3):854-861. doi:10.1016/j.ijrobp.2005.02.044

116. Nordsmark M, Bentzen SM, Overgaard J. Measurement of Human Tumour Oxygenation Status by a Polarographic Needle Electrode: An analysis of inter- and intratumour heterogeneity. *Acta Oncologica*. 1994;33(4):383-389. doi:10.3109/02841869409098433

117. Nozue M, Lee I, Yuan F, et al. Interlaboratory variation in oxygen tension measurement by Eppendorf "Histograph" and comparison with hypoxic marker. *Journal of Surgical Oncology*. 1997;66(1):30-38. doi:10.1002/(SICI)1096-9098(199709)66:1<30::AID-JSO7>3.0.CO;2-O

118. Collingridge DR, Young WK, Vojnovic B, et al. Measurement of Tumor Oxygenation: A Comparison between Polarographic Needle Electrodes and a Time-Resolved Luminescence-Based Optical Sensor. *Radiation Research*. 1997;147(3):329. doi:10.2307/3579340

119. Zhao D, Jiang L, Hahn EW, Mason RP. Tumor physiologic response to combretastatin A4 phosphate assessed by MRI. *International Journal of Radiation Oncology\*Biology\*Physics*. 2005;62(3):872-880. doi:10.1016/j.ijrobp.2005.03.009

120. Swartz HM, Williams BB, Zaki BI, et al. Clinical EPR: Unique Opportunities and Some Challenges. *Academic Radiology*. 2014;21(2):197-206. doi:10.1016/j.acra.2013.10.011

121. Khan N, Williams BB, Hou H, Li H, Swartz HM. Repetitive tissue pO2 measurements by electron paramagnetic resonance oximetry: current status and future potential for experimental and clinical studies. *Antioxid Redox Signal*. 2007;9(8):1169-1182. doi:10.1089/ars.2007.1635

122. Pogue BW, Feng J, LaRochelle EP, et al. Maps of in vivo oxygen pressure with submillimetre resolution and nanomolar sensitivity enabled by Cherenkov-excited luminescence scanned imaging. *Nature Biomedical Engineering*. 2018;2(4):254-264. doi:10.1038/s41551-018-0220-3

123. Cao X, Rao Allu S, Jiang S, et al. Tissue pO2 distributions in xenograft tumors dynamically imaged by Cherenkov-excited phosphorescence during fractionated radiation therapy. *Nature Communications*. 2020;11(1). doi:10.1038/s41467-020-14415-9

124. Bertolet A, Cortés-Giraldo MA, Carabe-Fernandez A. On the concepts of dose-mean lineal energy, unrestricted and restricted dose-averaged LET in proton therapy. *Physics in Medicine & Biology*. 2020;65(7):075011. doi:10.1088/1361-6560/ab730a

125. Cortés-Giraldo MA, Carabe A. A critical study of different Monte Carlo scoring methods of dose average linear-energy-transfer maps calculated in voxelized geometries irradiated with clinical proton beams. *Physics in Medicine and Biology*. 2015;60(7):2645-2669. doi:10.1088/0031-9155/60/7/2645

126. Sawakuchi GO, Ferreira FA, McFadden CH, et al. Nanoscale measurements of proton tracks using fluorescent nuclear track detectors: Nanoscale measurements of proton tracks. *Medical Physics*. 2016;43(5):2485-2490. doi:10.1118/1.4947128

127. Granville DA, Sahoo N, Sawakuchi GO. Simultaneous measurements of absorbed dose and linear energy transfer in therapeutic proton beams. *Physics in Medicine & Biology*. 2016;61(4):1765-1779. doi:10.1088/0031-9155/61/4/1765

128. Sawakuchi GO, Sahoo N, Gasparian PBR, et al. Determination of average LET of therapeutic proton beams using Al $_2$ O $_3$ :C optically stimulated luminescence (OSL) detectors. *Physics in Medicine and Biology*. 2010;55(17):4963-4976. doi:10.1088/0031-9155/55/17/006

129. Alsanea F, Therriault-Proulx F, Sawakuchi G, Beddar S. A real-time method to simultaneously measure linear energy transfer and dose for proton therapy using organic scintillators. *Medical Physics*. 2018;45(4):1782-1789. doi:10.1002/mp.12815

130. Terasawa K, Borak TB, Doke T, et al. The Response of the silicon-based Linear Energy Transfer Spectrometer (RRMD-III) to Protons from 1 to 70 MeV. *Japanese Journal of Applied Physics*. 2005;44(10):7608-7613. doi:10.1143/JJAP.44.7608

131. Zhang R, D'souza AV, Gunn JR, et al. Cherenkov-excited luminescence scanned imaging. *Optics Letters*. 2015;40(5):827. doi:10.1364/OL.40.000827

132. Brůža P, Lin H, Vinogradov SA, Jarvis LA, Gladstone DJ, Pogue BW. Light sheet luminescence imaging with Cherenkov excitation in thick scattering media. *Optics Letters*. 2016;41(13):2986. doi:10.1364/OL.41.002986

133. Kry SF, Alvarez P, Cygler JE, et al. AAPM TG 191: Clinical use of luminescent dosimeters: TLDs and OSLDs. *Medical Physics*. 2020;47(2):e19-e51. doi:10.1002/mp.13839

134. Fukumura A, Hoshino K, Takeshita M, Kanai T, Minohara S, Sudou M. SILICON DIODES AS DETECTORS IN RELATIVE DOSIMETRY OF HEAVY IONS. :8.

135. Bourhis J, Sozzi WJ, Jorge PG, et al. Treatment of a first patient with FLASH-radiotherapy. *Radiotherapy and Oncology*. 2019;139:18-22. doi:10.1016/j.radonc.2019.06.019

136. Venkatesulu BP, Sharma A, Pollard-Larkin JM, et al. Ultra high dose rate (35 Gy/sec) radiation does not spare the normal tissue in cardiac and splenic models of lymphopenia and gastrointestinal syndrome. *Sci Rep*. 2019;9(1):1-9. doi:10.1038/s41598-019-53562-y

137. Pappas EP, Zoros E, Zourari K, et al. PO-0774: Investigation of dose-rate dependence at an extensive range for PRESAGE radiochromic dosimeter. *Radiotherapy and Oncology*. 2017;123:S410. doi:10.1016/S0167-8140(17)31211-2

138. Sellakumar P, Samuel EJJ, Kumar DS. Dose-rate dependence of PAGAT polymer gel dosimeter evaluated using X-ray CT scanner. *J Phys: Conf Ser*. 2010;250:012074. doi:10.1088/1742-6596/250/1/012074